\documentclass[letter,aps,pre,showpacs,twocolumn,superscriptaddress,floatfix]{revtex4}
\usepackage{graphicx} 
\usepackage{amssymb}

\begin{document} 
  
\preprint{version 2} 

\title{Overcharging and charge reversal in the electrical 
double layer near the point of zero charge} 

\author{G. Iv\'an Guerrero-Garc\'{\i}a}
\affiliation{Instituto de F\'{\i}sica, 
Universidad Aut\'onoma de San Luis Potos\'{\i}, \\
\'Alvaro Obreg\'on 64, 78000 San Luis Potos\'{\i}, S.L.P., M\'exico} 
\affiliation{Programa de Ingenier\'{\i}a Molecular, Instituto Mexicano del 
Petr\'oleo,\\
Eje Central L\'azaro C\'ardenas Norte 152, 07730 M\'exico, D.F., M\'exico }

\author{Enrique Gonz\'alez-Tovar} 
\affiliation{Instituto de F\'{\i}sica, 
Universidad Aut\'onoma de San Luis Potos\'{\i}, \\
\'Alvaro Obreg\'on 64, 78000 San Luis Potos\'{\i}, S.L.P., M\'exico}

\author{Mart\'{\i}n Ch\'avez-P\'aez}
\affiliation{Instituto de F\'{\i}sica, 
Universidad Aut\'onoma de San Luis Potos\'{\i}, \\
\'Alvaro Obreg\'on 64, 78000 San Luis Potos\'{\i}, S.L.P., M\'exico}

\author{Marcelo Lozada-Cassou}
\affiliation{Programa de Ingenier\'{\i}a Molecular, Instituto Mexicano del 
Petr\'oleo,\\
Eje Central L\'azaro C\'ardenas Norte 152, 07730 M\'exico, D.F., M\'exico }

\date{\today}  
 
\begin{abstract} 

The ionic adsorption around a weakly charged spherical colloid, 
immersed in size-asymmetric 1:1 and 2:2 salts, is studied. We use the 
primitive model of an electrolyte to perform Monte Carlo simulations 
as well as theoretical calculations by means of the hypernetted chain/mean 
spherical approximation (HNC/MSA) and the unequal-radius modified 
Gouy-Chapman (URMGC) integral equations. Structural quantities such as 
the radial distribution functions, the integrated charge, and the mean 
electrostatic potential are reported. Our Monte Carlo ``experiments" 
evidence that near the point of zero charge 
the smallest ionic species is preferentially adsorbed 
onto the macroparticle, independently of the sign of the charge carried 
by this tiniest electrolytic component, giving rise to the appearance of
the phenomena of charge reversal and overcharging.   
Accordingly, charge reversal 
is observed when the macroion is slightly charged and the coions are larger 
than the counterions. In the opposite situation, i.e. if 
the counterions are larger than the coions, overcharging occurs (a feature 
originally predicted via integral equations in 
J.~Phys.~Chem.~B~{\bf108}, 7286 (2004), for the planar electrical 
double layer). In other words, {\it in this 
paper we present the first simulational data on 
overcharging}, showing that this novel effect 
surges, close to the point of zero charge, as 
a consequence of the ionic size asymmetry. We also find that the HNC/MSA 
theory captures well the charge reversal (CR) and overcharging (OC) phenomena 
exhibited by the computer ``experiments", 
especially as the macroion's charge increases. On the contrary, even if 
URMGC also displays CR and OC, its predictions do not 
compare favorably with the Monte Carlo results. Further, it is seen 
that the inclusion of hard-core correlations in Monte Carlo and HNC/MSA 
leads to spatial 
regions near the macroion's surface in which the integrated charge and/or 
the mean electrostatic potential can decrease when the colloidal charge 
is augmented and vice versa. These observations aware about the  
interpretation of 
electrophoretic mobility measurements using the standard 
Poisson-Boltzmann approximation beyond its validity region.

\end{abstract}  
 
  
\maketitle 
  
\section{Introduction}

It is widely known in physical-chemistry that a surface in contact with 
an electrolyte solution usually becomes charged and, thus, 
that the ions around the interface acquire a diffuse structure commonly 
denoted as the electrical double layer (EDL). One of the most successful 
early theories used to describe these systems in the dilute and/or weak 
electrostatic regimes is the classic Poisson-Boltzmann (PB) treatment, which 
is based on a point-ions representation of an 
electrolyte \cite{Ca02,Hiem01,But01}. Under 
this approach, it is an accepted fact that the counterions of a binary 
electrolyte are mostly adsorbed to the electrode when its surface charge 
density is increased; the coions, on the other hand, are pushed away from 
the region close to the charged surface \cite{Hiem01,Hu01}. 
As a result, in the PB 
picture, the main role of the EDL is to neutralize {\it monotonically} 
the surface charge as a function of the distance to the macroparticle, 
leading to a screened interaction between charged colloids in 
solution \cite{Gue01}. 
However, starting from the middle of the 
past century, 
an appreciable number of  experimental instances has been detected 
in which the effective charge of a colloid seems to reverse its 
sign (recent electrophoresis papers can be consulted in 
Refs. \cite{El01,Mar01,Qu01}). This 
singularity, due to an excess in the counterion's  compensation 
of the bare surface charge, 
is known as charge reversal (CR), 
and the too simplified PB formalism can not describe it since its 
occurrence involves strong ion-size correlations. It must be noted 
that, despite a lot of experimental research, it was only 
until recent times that direct measurements of CR were performed by 
Besteman et al. \cite{Bes01,Bes02,Hey01}, 
who employed atomic force microscopy techniques and a new 
electrophoresis capillarity apparatus to achieve laboratory 
conditions that were very difficult, or even impossible, to 
reach in traditional static and electrokinetic experiments. 

Through the years, the experimental advances have 
prompted the development 
of theoretical explanations for CR. In this way, the strong 
correlated liquid formalism \cite{Sh01} was conceived 
as an effort to comprehend and predict this kind of behavior. For 
macroparticles dissolved in trivalent electrolytes this scheme 
yields information consistent with observational data, but breaks down 
when applied to divalent solutions \cite{Hey01}. 
Alternatively, the theory of integral 
equations for liquids has proved to be a 
robust and reliable approach in this 
area of research 
\cite{Ca02,Ca01,Lo01,Vl01,Go01,Go02,Lo02,At01,Gr01}.  
For instance, it has been the main route to 
demonstrate that the ionic excluded volume constitutes, by itself, a 
physical mechanism with the ability to induce charge reversal \cite{Mes01}, 
although this does not mean, naturally, 
that the CR observed experimentally 
under very diverse conditions is due purely to such steric 
effects (an interesting discussion about the physical and/or 
chemical origins of CR is available in \cite{Ly01}).

Another fascinating phenomenon that has been theoretically 
predicted to occur in the EDL is overcharging (OC). OC appears 
when the coions are adsorbed to the electrified interface, despite 
the coulombic repulsion, {\it increasing} 
the native surface charge. This anomaly 
was first observed and defined for a mixture of macroions and a 
size-symmetric electrolyte in 
contact with a charged wall \cite{Ji01}. Notably, in such 
article the goal was to establish that 
the overcharging was prompted by the macroions, 
whereas here we will show that OC can 
be present even in a simpler system (without a wall) if 
ionic size asymmetry comes into play. In this context, it is worth 
to recall that the most basic way to 
incorporate {\it consistently} ionic 
size contributions into the EDL is by using the so-called primitive 
model (PM) of an electrolyte, wherein the ions are taken as hard 
spheres with point charges embedded at their centers. A large amount 
of work has been reported for the EDL using the PM in the 
special case of equal-sized ions (i.e. the restricted primitive model (RPM)), 
which implies a considerable simplification at the level of 
the model and calculations 
\cite{Ca01,Lo01,Vl01,Go01,Go02,Lo02,Mes01,Outh01,Lo03,Ye01,Des01,Go03,Yu01,Bh001,Ji02,Ji03,Goe01}. 
Notwithstanding, it is more realistic 
to expect EDL systems with different ionic sizes in their 
electrolytic species (due, for instance, to a distinct degree of 
hydration of the ions) and, 
thus, the PM should be preferred for a more faithful representation of 
the EDL.

Along these lines, the issue of the practical relevance of ionic
size asymmetry in the physical-chemistry of surfaces
has been recently revivified in an experimental
paper on the electrokinetics of uncharged colloids
by Dukhin et al. \cite{Du01}. In such investigation, the authors revisited the
idea that ``...a double layer might in fact exist, even when
there is no electric surface charge at all
(on the colloid), solely because of
the difference in cation and anion concentrations within the
interfacial water layer..." and provided 
a measurement technique and experimental data
supporting the existence
of the so-called zero surface charge double layer, 
a concept introduced
in a theoretical model by Dukhin himself and coworkers more 
than two decades ago \cite{Du02,De01,Du03}.
The relevant fact to our discussion is that, as proposed 
by Dukhin and other authors \cite{Jo01,Jo02,Co01,Ra01,Hie01,Man01}, 
such a difference in cation and anion concentrations, and
the concomitant charge separation in the proximity of an uncharged
colloid, can be attributed to the difference in the distances of closest
approach of counterions and coions to the surface
(as an alternative, or in addition, to the ``chemical" phenomenon
of specific adsorption).
Thence, the work by Dukhin et al. emphasizes the importance of
the ionic size asymmetry effect
in relevant phenomena occurring in real EDL systems, like the 
binding of simple inorganic electrolyte ions on colloidal 
substrates, and electrokinetics.
 
It is interesting that in the literature 
there are relatively few publications considering 
ionic size asymmetry~(see, for example, 
\cite{Gue01,Va01,Bh01,Sp01,Kh01,Ma01,Ou01,Gr02,Val01,Val02,Gi01,Gue02}), 
most of them dedicated to the planar geometry and 
only two reports for the spherical case \cite{Gue01,Gue02}. In our opinion, 
this apparent lack of interest in size-asymmetric 
electrolytes might arise from the long-standing belief in 
the dominance of counterions in the EDL; a fact 
foreseen and corroborated in a couple of pioneering 
papers on the modified Gouy-Chapman (MGC) theory 
for planar interfaces \cite{Va01,Bh01}. 
To be more explicit, the precise meaning 
of the dominance of counterions 
is that ``...at large potentials or charge 
densities, the coions are excluded from the vicinity of 
the electrode. Consequently, the counterions dominate and the 
double layer properties approach those of a symmetric 
electrolyte whose charge and diameter are equal to those of 
the counterions..." \cite{Bh01}. Noticeably, and even if 
this behavior was originally enunciated and verified 
{\it exclusively at the MGC level}, during the past years 
most of the modern studies of the EDL, that use 
theoretical approaches which surpass the punctual-ions PB treatment, 
have subscribed (or 
assumed without a rigorous proof) such counterion 
predominance in the primitive model 
\cite{Ca02,Kh01,Ma01,Val01,To02,Bo01,Yu01,Bh02}, 
a situation that has resulted in 
the mentioned scant attention to size-asymmetric EDLs. 
In contrast, very recently, some of the present authors have 
published an integral equation and simulational analysis of the 
size-asymmetric spherical EDL \cite{Gue01} where it has been evidenced that, 
for highly charged surfaces, counterions do not 
always dominate, i.e. that coions really 
matter in the double layer. At this point, it should be also noted that 
the establishment of the counterion dominance 
by Valleau, Bhuiyan and coworkers \cite{Va01,Bh01} was not really 
based on the {\it plain} Poisson-Boltzmann equation, or equivalently on 
a model of a genuine punctual electrolyte, since for these authors  
the ion-ion potential corresponds to that between point charges whereas 
for the ion-wall interaction a closest approach distance 
(hard-core or Stern correction) is added. In fact, the unequal radius 
modified Gouy-Chapman (URMGC) approach of Refs. \cite{Va01,Bh01} 
represents not only 
the inclusion, at the lowest order, of excluded volume contributions 
into the Gouy-Chapman theory (via the Stern modification),  
but also the first attempt to take into account the ionic size asymmetry 
through the use of different distances of closest approach for 
counterions and coions. From all the above discussed, 
the new integral equations results for 
a colloid in contact with a size-asymmetric PM 
electrolyte \cite{Gue01,Gue02} (in which hard-core  
contributions are consistently embodied in both 
the ion-surface and ion-ion interactions) 
imply that an incomplete consideration of excluded volume and size asymmetry 
contributions, like that of the URMGC theory, can lead to 
an inaccurate description of the double layer at high surface 
charges.

On the other hand, apart from the dominance of counterions
at large electric fields, URMGC has predicted other
notable phenomena in the EDL, this time
for low charged surfaces, such as the appearance of 
a potential of zero charge, oscillations in the
ionic density and mean electrostatic potential functions, 
and indeed 
charge reversal and overcharging \cite{JYu01,Mes02}. Lately, several 
theoretical 
and simulational investigations \cite{Val01,Gi01,To01}  
have pondered the planar EDL for a PM electrolyte 
in the low-charge regime 
and have corroborated the existence 
of the potential of zero charge and the non-monotonic behavior of 
the ionic radial 
distribution functions (RDFs) and potential profiles, previously 
seen in the ``semi-punctual" URMGC. Nevertheless, in the same 
reports it was also found that only those theoretical formalisms 
(e.g. the MPB5 and density functional theories) that 
fully include the hard-core and ionic size asymmetry 
effects succeeded in describing quantitatively the EDL near 
the point of zero charge \cite{Gi01,To01}, contrasting with URMGC 
that showed solely a limited 
success for 1:1 salts \cite{Val01}. In addition, 
and with respect to CR and OC, in a 2006 
paper \cite{JYu01} Yu et al. noticed, for the first time,  
the appearance of these ``anomalies" 
in an URMGC treatment of an electrolyte in a charged slit; 
a certainly intriguing fact given that, in the past,  
CR and OC had been observed only in 
theoretical analysis of primitive model EDLs \cite{Ji01,Mes01}. 
Therefore, at present, an exhaustive simulation study that confirms 
and characterizes the phenomena of charge reversal and 
overcharging in {\it slightly} charged PM EDLs, as well as 
an application of reliable theories in order to 
explain these striking features, is still lacking 
in the literature. 
Precisely, the main objectives of this communication are, in the 
first place,  
to supply fresh and comprehensive Monte Carlo (MC) data about 
the potential of zero charge, charge reversal, 
overcharging and the behavior of diverse structural 
properties of a low charged primitive model EDL in spherical geometry and, 
secondly, to present the comparison of such simulational information 
with the corresponding theoretical results of the HNC/MSA and 
URMGC integral equations, trying to assess the 
consequences of a consistent treatment of the excluded volume 
and ionic size asymmetry contributions in the spherical 
electrical double layer.

To investigate the static properties of the size-asymmetric 
spherical EDL in the weakly 
charged regime, with special focus on the CR and OC phenomena, 
we have produced simulation and HNC/MSA integral equation 
results for the radial distribution 
functions (RDFs) of 1:1 and 2:2 primitive model electrolytes bathing 
a spherical colloid under diverse conditions.
From the 
RDFs we extract the mean electrostatic potential 
and the charge profiles to identify 
the presence of charge reversal and overcharging and 
to examine their dependence on 
the ionic size asymmetry and other system parameters. 
All this structural information is contrasted 
with that corresponding to the semi-punctual URMGC theory. As it 
will be illustrated below, most of the computer 
``experiments" data are better paralleled by HNC/MSA than by URMGC. 
Additionally, we would like to note that, to the best of our knowledge, this 
work reports the first confirmation, via 
simulations, of charge reversal and overcharging in the EDL near the 
point of zero charge. This paper is organized as follows. In Section II 
we present the model system and the theoretical approaches.  
Section III contains the details of the numerical solution of the integral 
equations and of the Monte Carlo simulations. Section IV is devoted to the 
results and their discussion, and we close with a summary 
of relevant findings and conclusions in Section V.

\section{Model system and theory}

Our basic representation of the spherical electrical double layer 
(SEDL) is constituted by a rigid, charged
spherical  colloid of diameter $D_0$ and 
uniform surface charge density
$\sigma_0$, surrounded by a 
continuum solvent of dielectric constant $\epsilon$. The macroion is
in contact with two ionic species, which in 
the primitive model are treated as hard spheres
of diameters $D_i$ ($i=1,2$) with embedded point charges, 
$q_i$, at their centers. Note that  
$q_1q_2<0$. The interaction potential between the
particles in this model (i.e. macroion and electrolytic ions) is then given by

\begin{table}[htbf]

\begin{equation}
\label{umi}
U_{ij}(r)=\left\{
\begin{array}{ll}
\infty, & \textnormal{ for } r<D_{ij},\\
q_i q_j/(\epsilon r), & \textnormal{ for } r\geq D_{ij},
\end{array}
\right.
\end{equation}

\end{table}

\noindent where the subscripts $i$ and $j$ run from $0$ 
to $2$, $r$ is the center-to-center distance 
between two 
particles of types $i$ and $j$, 
$D_{ij}=(D_i+D_j)/2$, 
$q_{i}=z_{i}e$ is the charge of the species $i$ 
with valence $z_i$, $e$ is the protonic charge, 
and, for the spherical colloid, 
$q_{0} = z_0 e = 4 \pi (D_0/2)^2 \sigma_0$. The system as a whole 
is electroneutral, i.e.  $\sum_{j=1}^2 z_j \rho_j = 0$, where $\rho_j$ 
is the bulk number density of the electrolytic species $j$.

The Ornstein-Zernike equation for a multicomponent mixture 
of $S$ species is \cite{Mc01}

\begin{equation}
\label{ec1}
h_{ij}(r) = c_{ij}(r) +
\sum_{k=0}^{S-1} \rho_k \int 
h_{ik}(t)c_{kj}(|\,\vec r - \vec t\,| ) 
dV,
\end{equation}

\noindent such as $\rho_i$ is the bulk number density of each one
of the species in the system, $r=|\,\vec r\,|$ and $h_{ij}(r)$ 
are, respectively, the distance and the total 
correlation function between two particles of types $i$ and $j$, 
$g_{ij}(r) =h_{ij}(r) + 1$ is the radial distribution function, 
$c_{ij}(r)$ is the direct correlation
function, and $t=|\, \vec t\, |$ being the distance between two 
particles of types $i$ and $k$. 
This group of equations requires a second relation (or closure)
between the total and direct correlation functions. 
For charged systems, the
hypernetted chain (HNC) and the mean spherical approximation (MSA) 
closures are widely used \cite{Mc01,Han01}. The HNC and MSA relations, 
for $r\ge D_{lm}$, 
are given as:

\begin{equation}
\label{ec2}
c_{lm}(r) =  
-\beta U_{lm}(r) + h_{lm}(r)
- \ln [ h_{lm}(r) + 1 ],
\end{equation}

\noindent for HNC, and 

\begin{equation}
\label{ec3}
c_{lm}(r) =  -\beta U_{lm}(r),
\end{equation}

\noindent 
for MSA, where $\beta = 1/(k_B T)$ is the inverse of the thermal
energy. These expressions are complemented by the exact condition
$h_{lm}(r)=-1$, for  $r<D_{lm}$. 

Let us consider $S=3$ and that the species $0$ 
corresponds to macroions (thereinafter denoted equivalently 
by $M$) at infinite dilution, whereas 
the indices $1$ and $2$ are associated to a binary electrolyte. 
Then Eq. (\ref{ec1}) for the components 
$M (\equiv 0)$ and $j$ can be written as:

\begin{equation}
\label{ec4}
h_{Mj}(r) = c_{Mj}(r) +
\sum_{k=1}^2 \rho_k \int 
h_{Mk}(t)c_{kj}(|\,\vec r - \vec t \,| ) 
dV,
\end{equation}

\begin{center}
$j=1,2.$ 
\end{center}

\noindent Note that Eqs. (\ref{ec4}) are a complete set of integral equations 
for the SEDL. When Eq.~(\ref{ec2}) is employed in Eqs.~(\ref{ec4}) 
solely for $c_{Mj}(r)$, 
and the $c_{kj}(|\,\vec r - \vec t\,|)$ are 
approximated by the MSA 
analytical expressions for a bulk electrolyte \cite{Bl01,Bl02,Hir01}, 
the HNC/MSA integral equations are obtained. A detailed account of 
this HNC/MSA formalism can be consulted elsewhere \cite{Gue01} 
and will not be repeated here. Our election of the HNC/MSA theory 
to perform the present study is 
based on the fact that, for many years, it has been used successfully to 
investigate the RPM electrical double layer in many geometries (e.g. planar, 
cylindrical, and spherical) \cite{Lo01,Go01,Go02,Deg01,Deg02,Gue01}. 

The integral version of the URMGC theory in {\it spherical geometry} 
is easily deduced from the HNC/MSA formulation if 
$c_{kj}(|\, \vec r - \vec t \, |)=-\beta q_k q_j/(\epsilon |\, 
\vec r - \vec t \, |)$ is inserted in Eqs.~(\ref{ec4}), instead of the 
inter-ionic MSA direct correlation functions. It must 
be stressed that, in the original papers \cite{Va01,Bh01}, as well as 
in all the posterior treatments \cite{Val01}, URMGC has been solved, in 
differential form, strictly for planar interfaces, which means that 
the present study extends the classic URMGC planar theory to 
the spherical instance, continuing along the lines of our 
previous URMGC (and HNC/MSA) account of the SEDL \cite{Gue01}.

Once the $g_{Mj}(r)$ are available, either from a theory (e.g. 
HNC/MSA or URMGC) or from simulation, 
it is then possible to calculate various relevant functions, namely 
the local electrolyte charge density (LECD), 

\begin{equation}
\rho^*(r) = \sum_{j=1}^2 z_j \rho_j g_{Mj}(r) 4 \pi r^2, 
\label{rho_star}
\end{equation}

\noindent the total integrated charge (IC),

\begin{equation}
P(r) = z_M + \int_0^r \rho^*(t) dt, 
\label{ic}
\end{equation}

\noindent and the mean electrostatic potential (MEP),

\begin{equation}
\psi(r) =   \frac{e}{\epsilon} 
\int_r^\infty  \frac{P(t)}{t^2} dt. 
\label{pot2}
\end{equation}

\noindent These 
quantities are fundamental in our 
analysis of the properties of the SEDL. The LECD is 
a linear density that gives us 
information about the electrolytic charge (in units of $e$) 
inside a spherical shell of thickness $dr$ located at a distance $r$ 
from the center of the macroion. Besides, the integral 
over all the space of 
$-\rho^*(r)$ 
results in the valence of the macroion, $z_M$, 
as required by the electroneutrality condition. On the other hand, the IC is a measure 
of the net charge (in units of $e$) enclosed in a sphere of radius $r$ 
centered in the macroion. It equals $z_M$ for 
$D_0/2 \leq r \leq (D_0+D_1)/2$, if $D_1 < D_2$, and goes to zero as 
$r\rightarrow\infty$, again due to the electroneutrality restriction.
The IC also allows to compute the amount of charge adsorbed onto the 
macroparticle, i.e. the accumulated charge within the Helmholtz 
planes (see below for the definition of the Helmholtz planes), and, 
moreover, it 
detects charge reversal when $P(r)z_M < 0$, and 
overcharging if $P(r)z_M > 0$ and $|P(r)|>|z_M|$.
In addition, the MEP around the macroion is a central 
magnitude in colloid science because 
it determines, for instance, the regimes of 
stability/flocculation or the migration of macroparticles 
in a colloidal suspension \cite{Hu02}. 
As a matter of fact, the MEP at certain distance near the macroion's 
surface is usually identified with the well-known electrokinetic potential 
at the slipping plane (or zeta potential, $\zeta$) \cite{Hu01}.
The $\zeta$ potential is experimentally measurable in systems 
that display electrokinetic effects such as 
electrophoresis, electro-osmosis and 
streaming currents, and it has an ample use in 
the physico-chemical characterization, separation and/or fabrication  
of colloidal materials \cite{Hiem01,Fe01}. 

In particular, we will be interested
in the IC and the MEP in the neighborhood of the Stern layer. 
More especifically, the Stern layer is the free-of-ions space 
next to a macroion that ends at the Helmholtz plane. 
The Helmholtz plane (or, more properly, the Helmholtz surface) is the
geometrical place associated to the colloid-ion closest approach
distance \cite{Hu01,Hu02}. In size-symmetric electrolytes only one Helmholtz 
plane can be identified. In our model, however, the size asymmetry 
between the ions allows us to define an inner
Helmholtz ``plane" (IHP) and an outer Helmholtz ``plane" (OHP) (note the  
conventional usage of the word ``plane"). 
The IHP is
specified by the closest approach distance of the smallest ionic
component to the colloidal surface (i.e., by $(D_{0}+D_{1})/2$, if
$D_{1}<D_{2}$), whereas the OHP is determined by the corresponding
distance of closest approach for the largest species (i.e., by
$(D_{0}+D_{2})/2$, if $D_{1}<D_{2}$). Therefore, for the primitive 
model EDL, if $D_1 < D_2$, the Stern layer is the region 
where $D_{0}/2\leq r < (D_{0}+D_{1})/2$, 
and the Helmholtz zone corresponds to 
$(D_{0}+D_{1})/2\leq r \leq (D_{0}+D_{2})/2$. 
Obviously, for size-symmetric salts the
IHP and the OHP coincide and the standard notions of the 
Helmholtz plane and Stern layer are recovered. 
In the general PM case, and provided that $D_1<D_2$, 
when the MEP is evaluated 
at $r=(D_0+D_1)/2$, Eq. (\ref{pot2}) gives the MEP at the IHP, which 
we denote as $\psi_{IHP}$. On the other hand, if Eq. (\ref{pot2}) 
is calculated at $r=(D_0+D_2)/2$, the MEP at the OHP, 
$\psi_{OHP}$, is obtained.

\section{Numerical details}

A size-asymmetric 1:1 or 2:2 electrolyte with a ratio between 
ionic diameters $D_-/D_+=2$, bathing a charged macroparticle of diameter 
$D_M=D_0=20$ \AA~ and valence $z_M$, was considered in all  
the calculations reported. Specifically, the diameters of the
positive and negative species were $D_+=4.25$~\AA~ and $D_-=8.5$~\AA,
respectively. In other words, for definitiveness, the cations have 
been chosen as the smallest species 
of the binary electrolyte, i.e., $D_+=D_1$ 
and $D_-=D_2$, with $D_1<D_2$. 
Note also that the sign of $z_M$ defines which one of the ionic
species is the coion or counterion.

To establish the primitive and semi-punctual models 
employed in our simulational and theoretical approaches, let 
us introduce 
the macroion-ion contact distances, 
$d_{M+}$ and $d_{M-}$, 
given by

\begin{table}[htbf]

\begin{equation}
\label{max_col_ion_sim}
d_{Ml}=\left\{
\begin{array}{ll}
(D_M+D_{+})/2,&\textnormal{ for }l=+ \textnormal {,} \textnormal { in simulation,} \\
& \textnormal { HNC/MSA and URMGC,}  \\
(D_M+D_{-})/2,&\textnormal{ for }l=- \textnormal {,} \textnormal { in simulation,} \\
& \textnormal { HNC/MSA and URMGC.} 
\end{array}
\right.
\end{equation}

\end{table}

\noindent It must be remembered that here the macroions, cations and 
anions correspond to the indices $0$, $1$ and $2$, respectively. 

Complementarily, the ion-ion contact distances, $d_{+\,+}$, 
$d_{-\,-}$, and $d_{+\,-}(=d_{-\,+})$, are 

\begin{table}[htbf]

\begin{equation}
\label{max_ion_ion_sim}
d_{ij}=\left\{
\begin{array}{llll}
D_{+},&\textnormal{ for }i=j=+ \textnormal {, in simulation} &\\
& \textnormal { and HNC/MSA,} & \\ 
D_{-},&\textnormal{ for }i=j=- \textnormal {, in simulation}&\\ 
& \textnormal {  and HNC/MSA,} & \\

(D_{+}+D_{-})/2,&\textnormal{ for }i=+ \textnormal { and }j=- \textnormal {, in simulation}& \\
& \textnormal {  and HNC/MSA,} \\

0,&\textnormal{ for any }i \textnormal { and } j \textnormal {, in URMGC.} &\\

\end{array}
\right.
\end{equation}

\end{table} 

\noindent 
The dielectric constant and temperature considered in all the cases were
$\epsilon=78.5$ and $T=298\,K$. 

The URMGC and HNC/MSA theories were numerically solved by means of a 
Picard iteration scheme, which, in the past, has been thrivingly 
employed in a number 
of studies of the EDL in various geometries via integral equations and 
density functional theories \cite{Gue01,Yu01,Goe01}.

Simulations were performed in a cubic box with the 
usual periodic boundary conditions in the canonical ensemble. 
The ionic species satisfied the electroneutrality condition: 
$z_M + N_+ z_+ + N_- z_- + N_c z_c= 0$, where $z_M$ is the valence 
of the macroion, $N_+$, $z_+$ and  $N_-$, $z_-$ are the number of ions 
and the valence of the positive and negative 
electrolytic species, respectively, 
and $N_c$ and $z_c$ are the number and the valence of the counterions that 
balance the colloidal charge.   
In order to accomplish consistent comparisons with the theory, 
the absolute  value of the  valence of such counterions was $|z_c|=1$ 
for the 1:1 electrolyte and $|z_c|=2$  for  the 2:2 salt. The macroion 
was fixed in the center of a simulation box of  
length $L$ and, in order to avoid border effects, the extension of the cell 
was enlarged until the integrated charge showed 
clearly a plateau of zero charge 
far from the macroion. The runs were done for $\approx$ 2000 particles 
for the monovalent salt and $\approx$ 1000  ions for the divalent electrolyte. The long-range 
interactions  were taken into account by using the Ewald sum 
technique with conducting  boundary conditions \cite{Al01,Fr01}. 
The associated damping parameter was  $\alpha= 5/L$ and 
the $\vec{k}$-vectors employed to compute the reciprocal space contribution 
satisfied the condition $|\vec{k}| \leq 5$. Monte Carlo runs of charged systems were 
performed with $5 \times  10^4$  equilibration MC cycles and from $6 \times 10^5$ to 
$1.8 \times 10^6$ MC cycles were practised for the production stage.
The radial  distribution 
functions were calculated using standard bin procedures \cite{Al01,Fr01}, 
whereas the integrated charge and the electrostatic potential were 
obtained  by using Eqs. (\ref{ic}) and (\ref{pot2}), respectively.

\section{RESULTS AND DISCUSSION}

In what follows, our results are discussed chiefly in terms of the 
Monte Carlo (MC) simulations data. Appropriate comparisons with the 
HNC/MSA and URMGC formalisms are presented such that the accuracy 
of the theoretical predictions  can be assessed. 

\subsection{Monovalent size-asymmetric electrolytes}

The structure of the EDL is the result of an entropic and energetic
competition. In the proximities of the point of zero charge 
the entropy is expected to be important,
whereas for highly charged macroparticles the EDL is expected to exhibit
strong coulombic correlations. Thus, in order to understand
the behavior of the ionic atmosphere next to a barely charged colloid 
in contact
with a size-asymmetric electrolyte, 
we will first present a comparison between the surrounding 
distribution of a univalent salt and that of a  mixture of hard spheres
having the same radii and concentration. This will illustrate how
the structure of charged systems deviate from the neutral situation. 

\begin{figure}[h]
\begin{center}
\includegraphics[width=8.0cm]{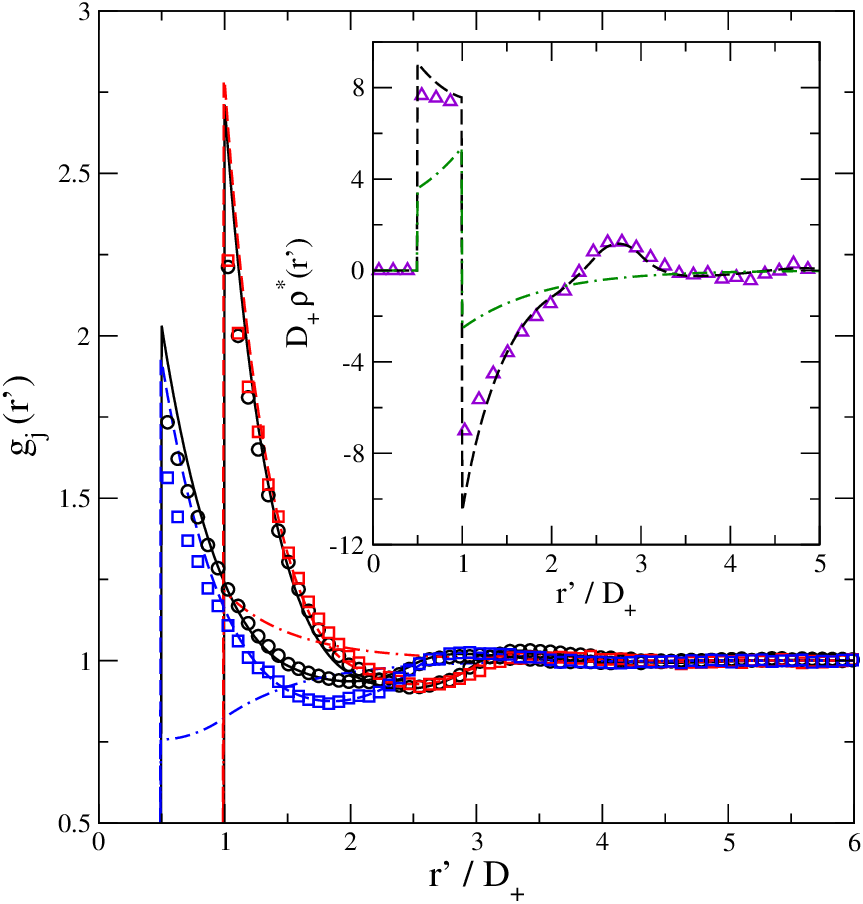}
\caption{Radial distribution functions of a size-asymmetric binary mixture of
  hard spheres and a size-asymmetric 1:1 electrolyte, both at a 1 M
  concentration, around an uncharged ($z_M=0$) macroparticle (in the main panel). 
  Thereinafter, the diameters of the ionic species and colloid are $D_{+}=4.25$~\AA, 
  $D_{-}=8.5$~\AA, and $D_{M}=20$~\AA, respectively. The small hard spheres 
  have the same diameter as that of the cations, $D_{1}=D_{+}$, and the large hard spheres 
  diameter is the same as that of the anions, $D_{2}=D_{-}$. 
  The open circles 
  correspond to Monte Carlo (MC) simulations of the size-asymmetric 
  mixture of hard spheres and the open squares represent the MC data 
  of the size-asymmetric monovalent electrolyte. The
  solid lines are associated to the asymmetric mixture of hard spheres in
  the HNC/MSA approach. The dashed and dot-dashed lines 
  are the HNC/MSA and URMGC results for the 1:1 size-asymmetric 
  salt, respectively. In the inset we depict the corresponding local electrolyte 
  charge density for the 1:1 electrolyte case. Open triangles, 
  and dashed and dot-dashed lines correspond to simulations, and the HNC/MSA 
  and URMGC theories, respectively. Here and in the rest of the figures, the 
  distance $r$' is measured from the colloidal surface.  
}
\label{fig_c4_pm_z0_hs}
\end{center}
\end{figure}

Let us consider initially the ionic distribution around an uncharged 
macrosphere ($z_M=0$). Fig. \ref{fig_c4_pm_z0_hs} (main panel) 
displays the radial
distribution functions (RDFs) of two systems; in one case the 
ionic species represent
a 1:1, 1 M salt, and in the other the ions are
uncharged, forming a pure hard-sphere assembly. 
We note here that, in the former system, the EDL is the result of 
both the entropic and energetic contributions; in contrast, 
the structure of the latter instance is only driven by entropy because 
all the interactions are of the excluded volume type.
Nevertheless, since we are working in the zero colloidal charge regime, 
we expect that in the first case the excluded volume interactions must play a 
determinant role in the resulting properties of the double layer.
The direct comparison of the simulational and theoretical RDFs  
of the charged and uncharged spheres around the neutral macroion
is presented in the main panel of 
Fig. \ref{fig_c4_pm_z0_hs}. 
An inspection of the Monte Carlo data shows that the 
structure of the two mixtures (of ions and of the hard spheres) is rather similar  
for $r'/D_+>2.5$ 
(where $r'$ is the distance measured from the colloidal surface), 
indicating the existence of weak charge correlations at these distances. 
Differences, however, are noticeable at smaller distances, 
particularly for the tiniest species. We observe that the addition 
of charge to the ionic species slightly increases the concentration of the 
larger species and decreases the concentration of the smaller species, 
especially at the contact distances. As it is shown later, this dewetting of 
the surface augments with the valence of the ions. 
Such comportment of the RDFs could lead eventually 
to significant changes in the thermophysical properties 
of the charged systems since, for instance, the pressure 
depends directly on the contact values of the pair 
correlation functions \cite{Han01}. 
We must point out that these are rather concentrated systems, 
with an ionic volume 
fraction of $\phi\approx 0.217$. That is the reason why the contact peaks 
of the neutral system are so high. 
Hence, at these volume fractions, the charge correlations 
in the neighborhood of the macroparticle are masked by the strong 
steric contributions. Notwithstanding, we remark that, for barely  
charged systems, the absolute values of the ionic RDFs close to 
the surface and the extent of their deviations with respect to the pure 
hard-sphere mixture will rule the behavior of the IC and MEP and, 
as it is evidenced below, will be crucial to the degree of 
appearance of CR and OC. 
 
The adequacy of a theoretical description of 
the present EDL systems will depend on its  
ability to capture correctly the electrostatic 
and steric correlations close to the macroparticle.
In this regard, the integral equations results portrayed in  
Fig. \ref{fig_c4_pm_z0_hs} illustrate 
that HNC/MSA follows closely the trends of the simulations,
with quantitative discrepancies near the colloid  
where this scheme overestimates the RDFs. 
In contrast, the URMCG data 
exhibit very different tendencies from those of the simulations. 
Especially noticeable is the pronounced separation between the 
URMGC radial distribution function of small ions and those from 
HNC/MSA and Monte Carlo, as well as the very low values 
of the URMGC normalized density of cations in the zone 
comprised by the Helmholtz planes. In fact, such 
exaggerated absence of small ions in URMGC 
will be of consequence in our 
later analysis of charge reversal and overcharging. 
Also, and contrary to the 
MC data, the URMCG theory predicts $g(r)$s that are monotonic beyond the OHP,  
a well documented characteristic of point-ions 
theories \cite{Hu01,Hu02,Hiem01}. 
From the previous discussion, 
it is therefore expected that the ensuing properties of the 
EDL extracted from the 
HNC/MSA and URMGC ionic profiles should present important 
discrepancies, with HNC/MSA
excelling in the comparison with Monte Carlo.

To better visualize the distribution of charge around the neutral 
macroparticle, 
let us focus on the function $\rho^*(r)$. 
The analysis of the inset of 
Fig. \ref{fig_c4_pm_z0_hs} shows that the Monte Carlo LECD 
around the colloid exhibits a structure with 
alternating domains of positive and negative local charge.
Three intervals are clearly seen in the scale of the graph. 
The first domain is positive and encompasses the region between the
Helmholtz planes. 
A second space, where $\rho^*(r)$ is negative, 
spans from the OHP up to $r'/D_+\approx 2.4$, and a third interval, where
$\rho^*(r)>0$, is located at $2.4 \lesssim r'/D_+\lesssim 3.4$.
In the first domain the local charge is positive because the larger ionic 
species (the anions) is completely excluded from this region. 
Beyond the OHP $\rho^*(r')$ drops sharply to a large negative value,
due to the presence of a compact contact layer of anions at the OHP. The
charge remains negative in this second region until the concentration of positive and 
negative ions equalize, namely at the first crossing point of the
corresponding radial distribution functions. Clearly in the third zone 
the population of cations dominates, leading to
the positive charge observed in the graph. This pattern of alternating 
domains continues as one goes away from the surface of the macroion,
although they are not distinguishable in the scale of the figure. 
Again, we should note that HNC/MSA parallels very nicely the MC simulations, 
whereas URMGC predicts a monotonic, asymptotic increase of 
the local charge
outside the Helmholtz planes. Thus, for the last theory,
the charge outside the Helmholtz planes, at any distance from the
macroion, is always dominated by the anions, which departs evidently from the
predictions of Monte Carlo and HNC/MSA.

\begin{figure}[h]
\begin{center}
\includegraphics[width=8.0cm]{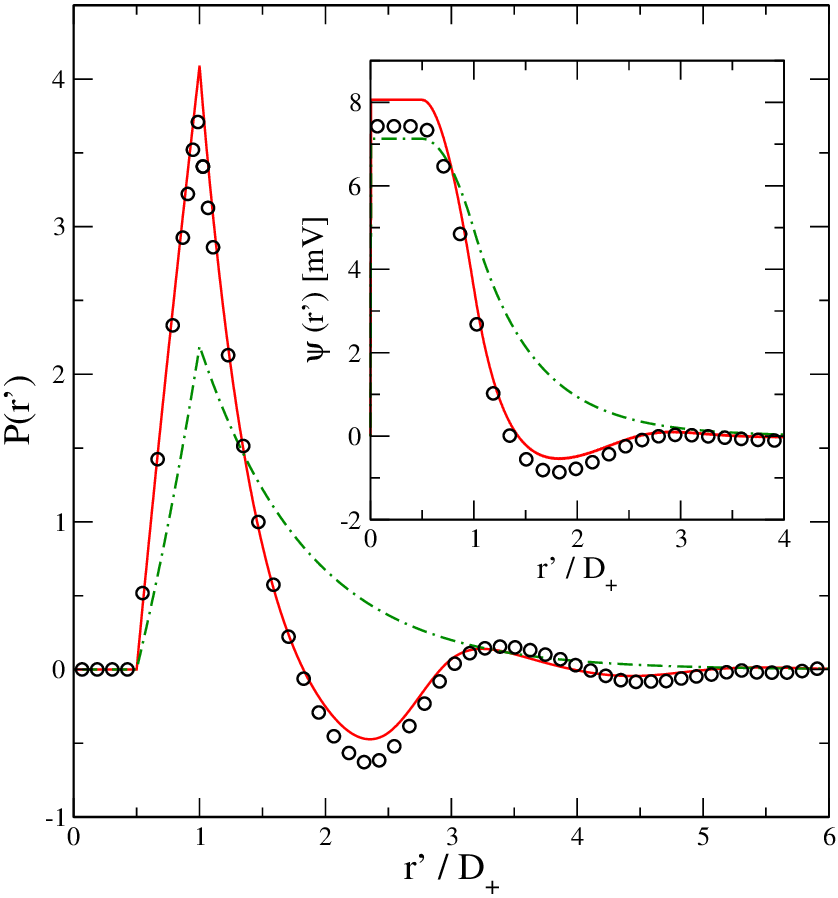}
\caption{Integrated charge (main panel) and mean electrostatic potential (inset) of a 
  size-asymmetric 1:1, 1 M electrolyte around an uncharged ($z_M=0$)
  macroparticle. 
  The open circles, solid and
  dot-dashed lines correspond to the MC, HNC/MSA and URMGC data,
  respectively.  
}
\label{fig_c4_pm_z0_qr_vr}
\end{center}
\end{figure}

Complying with the behavior of the local charge, and given 
that $\frac {dP(r')}{dr'}=\rho^*(r')$ (see Eq.(\ref{ic})), the 
$P(r')$ variates as it is observed in Fig. 
\ref{fig_c4_pm_z0_qr_vr} (main panel). Interestingly, we note that all 
the simulational and theoretical IC curves present 
the adsorption of a layer of charge very close to 
the macroion's surface; such layer begins at the IHP and
reaches its maximum value of charge at the OHP. The amount of charge in 
the Helmholtz zone 
is significant ($\approx 3.8e, 4.1e$ and $2.2e$ for MC, HNC/MSA and URMGC, respectively) and, 
in turn, gives rise to the formation of a double layer beyond 
the OHP (see Fig. \ref{fig_c4_pm_z0_hs}). 
The existence of this zero surface charge double layer (ZSC-DL) in our 
calculations is regarding since confirms the ideas originally proposed by 
Dukhin et al.  \cite{Du01} and, in addition, agrees with previous studies 
of the ZSC-DL in planar geometry via Monte Carlo simulations \cite{Val01}, 
and the URMGC and density functional theories \cite{Va01,Gi01}. 
In the MC and HNC/MSA cases, after its maximum, 
the IC decreases until changing sign and, 
subsequently, exhibits a region of negative values 
for $1.8 \lesssim r'/D_+ \lesssim 3.0$.
Farther than $r'/D_+ = 3.0$ the accumulated charge fluctuates
around zero and, finally, the electroneutrality condition
is obtained when $r' \to\infty$. Contrarily, the $P(r')$ of URMGC 
goes uniformly to zero.

The information presented so far clearly evince that, at the point of 
zero charge, a very
simple 1:1 salt in contact with a macroparticle already 
displays highly nonlinear 
effects such as charge adsorption (i.e. a 
ZSC-DL) or the reversion in 
the sign of the IC. These phenomena
are due to both the finite ionic size and 
to the asymmetry between the electrolytic species of the model.
Since HNC/MSA fully incorporates such conditions, it is able to
reproduce all the characteristics observed in the simulations, 
even at a quantitative 
level. In contrast, URMGC does not embody completely  
the ionic size correlations, just the non-zero contact distances 
between the colloid and ions. 
This is enough to capture the adsorption of charge, but
not the additional traits of the accumulated charge 
at intermediate and large $r'$. 
In particular, URMGC fails to
detect the sign reversal in $P(r')$, 
predicting instead a monotonic neutralization of the effective charge
adsorbed inside the Helmholtz planes. 
In other words, the partial inclusion of the ionic size 
and size asymmetry contributions in URMGC has the severe 
inconvenience that, in this semi-punctual approach, the 
occurrence of steric-related peculiarities, such as the 
ZSC-DL, charge reversal and the oscillation of the RDFs and IC, 
is restricted exclusively to the region between the 
Helmholtz planes. This fact, to which we will refer to as 
{\it the localization of the size and asymmetric effects in URMGC},  
will be a recurrent issue in our posterior 
discussions of the MEP and overcharging. 

The mean electrostatic potential 
as a function of the distance to the uncharged macroparticle is plotted in the 
inset of Fig. \ref{fig_c4_pm_z0_qr_vr}. 
The first thing worth to be noticed is that, 
despite a zero charge on the colloid, 
the MC, HNC/MSA and URMGC mean electrostatic potentials at 
the IHP are positive.  
This potential of zero charge (PZC) has been 
largely recognized as a direct 
consequence of the ionic size asymmetry in the 
EDL since the initial papers by Valleau and others \cite{Va01,Bh01}  
and, in the mean time, it has received great attention 
in diverse simulational and theoretical accounts of 
the planar double layer \cite{Gr02,Val01,Gi01}.
Evidently, this PZC 
is due to the dominant population of cations close to the macroion.
Far from the surface of the macroion the MEP of MC and of HNC/MSA tend to zero, 
but for intermediate distances they have a series of minima and maxima
of alternating sign. In particular, the first minimum defines the region
where the MEP reversal  
is stronger. Now, if the wonted association between the electrokinetic 
and mean electrostatic potentials is invoked \cite{Hu01},  
the mentioned comportment of the MEP suggests that 
a macroparticle could experience electrophoresis, even if 
it is uncharged. Thence, for small cations and depending on the precise localization of the
slipping surface \cite{Hu02}, the neutral colloid should move in the direction of the applied 
field if this surface is somewhere in between the Helmholtz planes,
or backwards if the shear boundary 
is around the first minimum. 
It is generally accepted that the slipping or $\zeta$-plane is very close 
to the surface of the macroions \cite{Hu02}, therefore we expect the first 
scenario to be more plausible. 
Note that UMRGC foresees that the macroions should flow always 
in the direction of the external electric field.

\begin{figure}[h]
\begin{center}
\includegraphics[width=8.0cm]{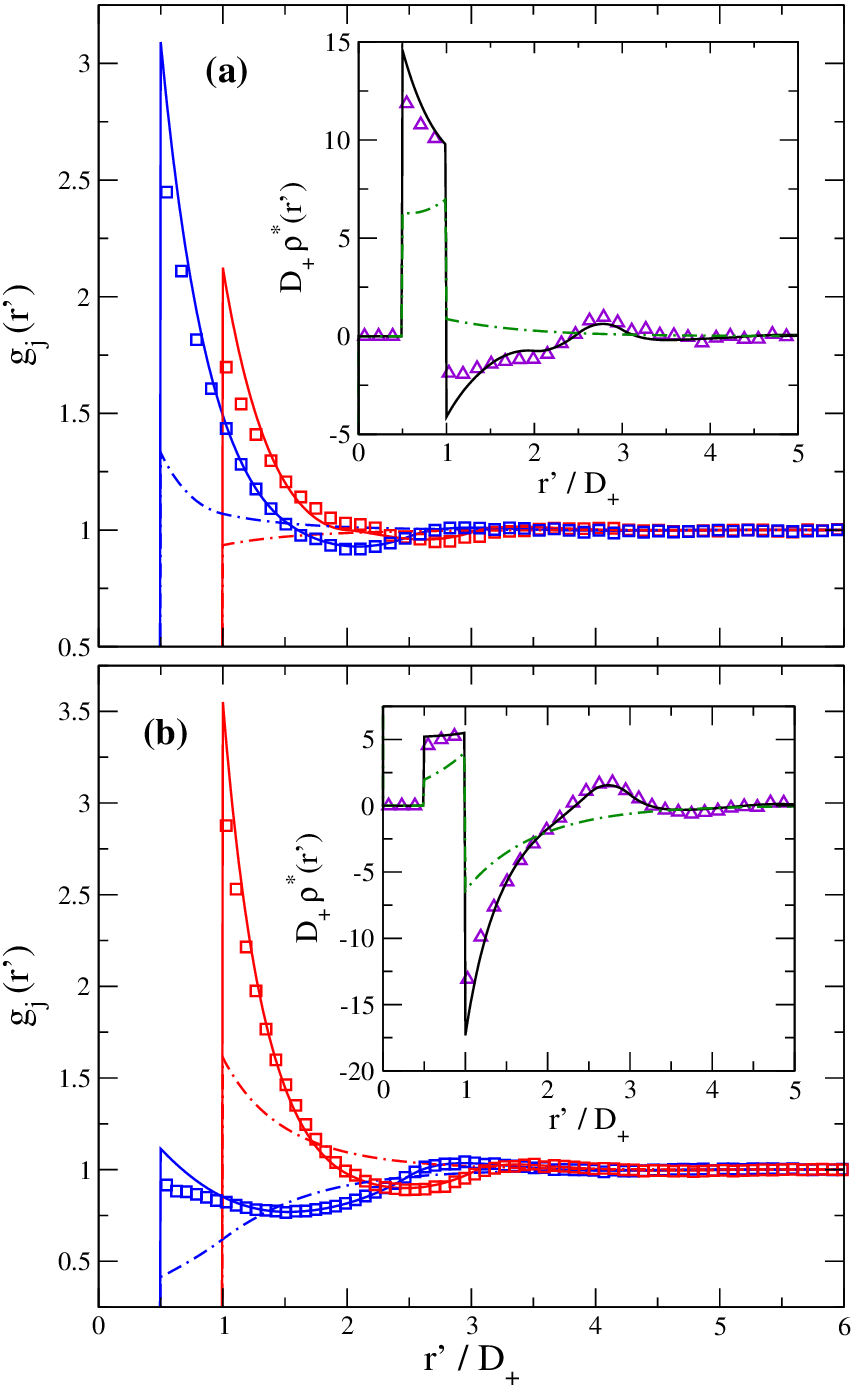}
\caption{Radial distribution functions of a size-asymmetric 1:1, 1 M
  electrolyte,  around a charged macroion (main panel). 
  In Fig. \ref{fig_z4_c2_c4}(a) $z_M=-4$, and in Fig. \ref{fig_z4_c2_c4}(b) $z_M=4$.  
  The open squares, solid and dot-dashed lines represent the MC, 
  HNC/MSA and URMGC results, respectively. In the insets the corresponding 
  local electrolyte charge densities are portrayed. Open triangles, 
  and solid and dot-dashed lines are associated to simulations, and the HNC/MSA 
  and URMGC theories, respectively.     
}
\label{fig_z4_c2_c4}
\end{center}
\end{figure}

So far we have examined systems at 
the point of zero charge, where no electrostatic 
correlations between the macroparticles and the
surrounding ions were considered. In spite of that, the 
colloid-ion entropic contributions 
and the inter-ionic correlations led to interesting phenomena. By weakly
charging the macroions, conspicuous effects such as charge reversal and
overcharging now arise, as it is seen in the remaining of this section.
Then, let us evaluate two situations in which the valence of 
the macroion is $z_M=-4$ and $z_M=4$ (i.e. surface charge densities  
$\sigma = \pm 0.051$ C/m$^2$), immersed in the same 1:1 electrolyte 
as before. Note that by virtue of $z_M$ the role of anions 
and cations as counterions 
or coions is interchanged. 
Fig. \ref{fig_z4_c2_c4} includes the radial structure (main panel) 
and the local charge (insets) of the electrolyte 
around the macroion for the two values of $z_M$. 
Fig. \ref{fig_z4_c2_c4}a contains the case
in which the counterions are smaller than the coions ($z_M=-4$), 
whereas the case of larger counterions ($z_M=4$) is reported in 
Fig. \ref{fig_z4_c2_c4}b. Fig. \ref{fig_z4_c2_c4}a reveals that, 
compared to the RDFs of a neutral macroparticle in 
Fig. \ref{fig_c4_pm_z0_hs}, 
the presence of the counterions in MC and HNC/MSA is greater when 
the surface is negatively charged, as evidenced by the increased 
contact peak, whereas the concentration of coions diminishes. 
This fact stresses 
the relevance of charge correlations (induced by $z_M$) on the ordering 
of the ions around the macroparticle. Clearly, the RDFs of URMGC disagree with 
MC and HNC/MSA. On the other hand, for 
$z_M=4$ (counterions larger than coions, Fig. \ref{fig_z4_c2_c4}b), 
we see a dramatic decrement in the contact peak of the smaller ions 
(coions) and an augment in that of the larger ions 
(counterions), which is adequately reproduced by 
HNC/MSA and not by URMGC. Since we are dealing with charged systems, 
size and charge correlations 
are into play. As a result, these two cases produce LECD functions 
that look very different 
(see the insets of 
Fig. \ref{fig_z4_c2_c4}). In addition, both insets show that the local charge 
profiles of simulation and HNC/MSA exhibit a sequence of positive and negative regions. 
Quantitatively, however, the first positive zone 
is much more pronounced for $z_M=-4$, whereas for $z_M=4$ the second
and third domains are much more important. In particular, it is worth to 
notice that, for simulation and theories, the first region is positive in the two cases, even 
if the charge of the macroion is positive, which indicates that, 
for $z_M > 0$, this 
first layer, rather than to screen the bare charge of the macroion, it 
emphasizes such native charge, which is a 
remarkable effect induced by the smaller 
size of the coions. 

\begin{figure}[h]
\begin{center}
\includegraphics[width=8.0cm]{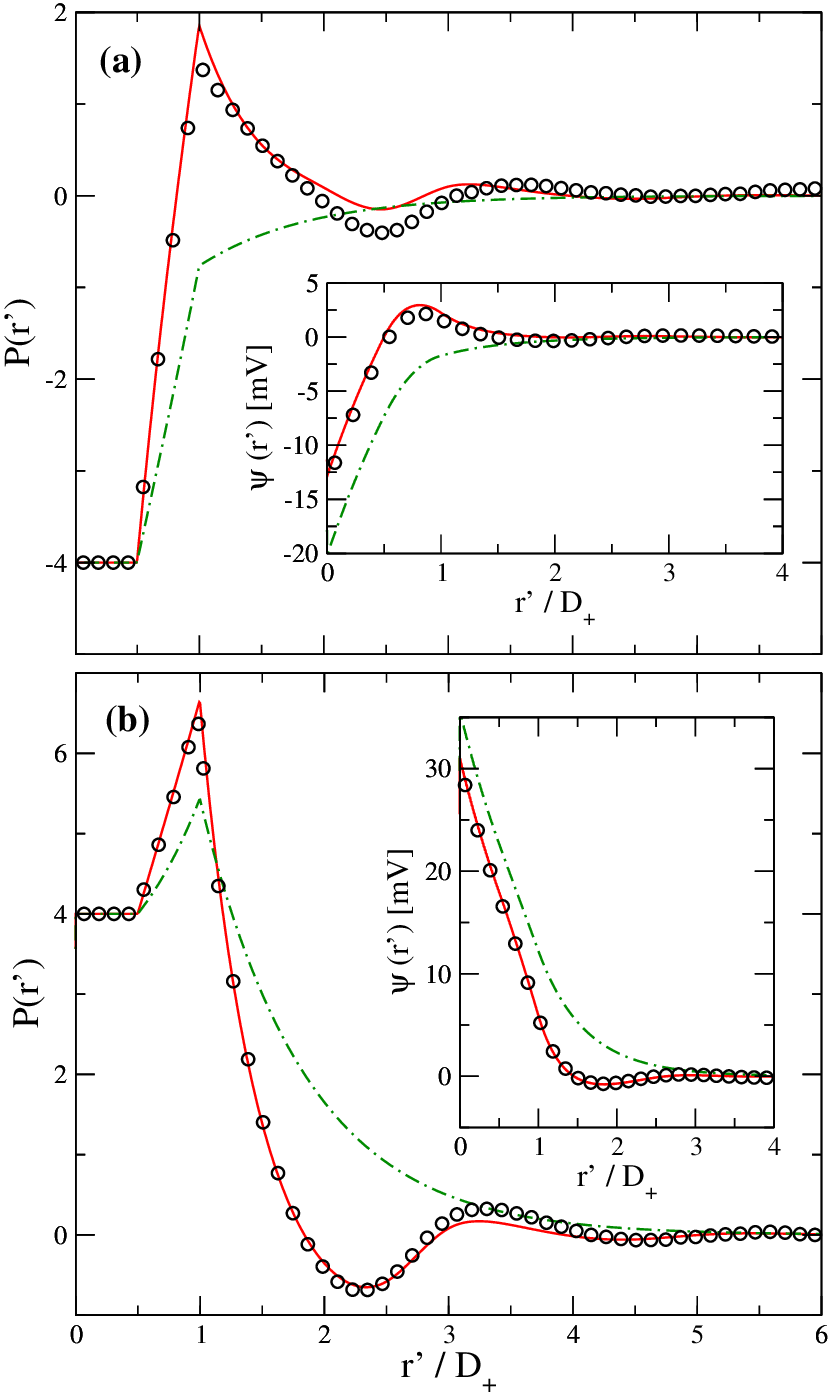}
\caption{Integrated charge (main panel) and mean electrostatic potential (inset) of a 
  size-asymmetric 1:1, 1 M electrolyte around a charged 
  macroion. 
  In Fig. \ref{fig_z4_c2_c4_qr_vr}(a) $z_M=-4$, and in Fig. \ref{fig_z4_c2_c4_qr_vr}(b)
  $z_M=4$.  
  The open circles, solid and
  dot-dashed lines correspond to the MC, HNC/MSA and URMGC data,
  respectively. 
}
\label{fig_z4_c2_c4_qr_vr}
\end{center}
\end{figure}

Further details can be grasped by looking at the
$P(r')$ curves, 
as they are presented
in Fig. \ref{fig_z4_c2_c4_qr_vr}. 
For $z_M=-4$ (Fig. \ref{fig_z4_c2_c4_qr_vr}a) 
we find that the Monte Carlo IC increases almost 
linearly inside the Helmholtz 
planes, inverting its original sign ({\it i.e. experiencing 
charge reversal}) 
and reaching a maximum of $P_{max}(r') \approx 1.3$ at the OHP. 
After the  OHP, $P(r')$ acquires an
fluctuating behavior, where subsequent charge inversions can be 
appreciated. The inset of the figure indicates that the mean 
electrostatic potential is also oscillatory, 
with a maximum inversion inside the Helmholtz zone. 
The level 
of accuracy of the theories can be readily estimated from the figure 
and its inset. Particularly, we see that, differently from 
HNC/MSA, URMGC is unable to describe any
reversal of the accumulated charge or the mean electrostatic potential 
throughout all the space. 
This represents an extreme manifestation of the 
so-called localization 
of the size and asymmetric effects in URMGC.

The functions $P(r')$ and $\psi(r')$ for $z_M=4$ are 
portrayed in Fig. \ref{fig_z4_c2_c4_qr_vr}b and its inset. 
Notably, the Monte Carlo $P(r')$ {\it increases} 
in between the Helmholtz 
planes, revealing {\it an adsorption of charge of the same sign 
as that of the macroion}, which in turn ``augments" the native 
macroion charge up to a maximum value of 
$P_{max}(r') \approx 6.3$ at the OHP. 
This striking event is referred to as {\it overcharging}. Overcharging 
(OC) was predicted theoretically 
by Jim\'enez-\'Angeles and Lozada-Cassou \cite{Ji01} for a 
charged wall in contact with a mixture of macroions and a fully-symmetric 
salt. To this date, this peculiarity had not been confirmed through
simulations or experiments. Hence, our results
provide the first simulational evidence of OC 
in a very simple model, where such ``anomaly"
is a direct consequence of the ionic size asymmetry when charge 
correlations are not too strong, i.e. close to the point 
of zero charge. In this
sense, overcharging appears when size correlations dominate
over charge correlations near the point of zero charge. This implies, therefore, 
that OC is also expected in other 
geometries, whenever ionic size asymmetry is present. 
We would like to point out that the results presented here are in line with 
our previous reports \cite{Gue01,Gue02}, where it has been established that 
the properties of the electric double layer depend on 
the two species of a binary electrolyte, and not only on the counterions 
as it has been widely accepted. On the other hand, 
regarding the performance of the 
theoretical
schemes we are working with, we realize that 
URMGC is capable to describe the 
overcharging, but not the oscillations and the charge reversal 
observed in the MC data, nor the inversion of the 
corresponding MEP. 
This fact is a typical expression of the URMGC restraint 
of all the excluded volume and size asymmetry phenomena 
to occur exclusively in the Helmholtz zone. 
Contrastingly, HNC/MSA collates very well 
with the simulational data, describing correctly all 
these characteristics.

\begin{figure}[h]
\begin{center}
\includegraphics[width=8.0cm]{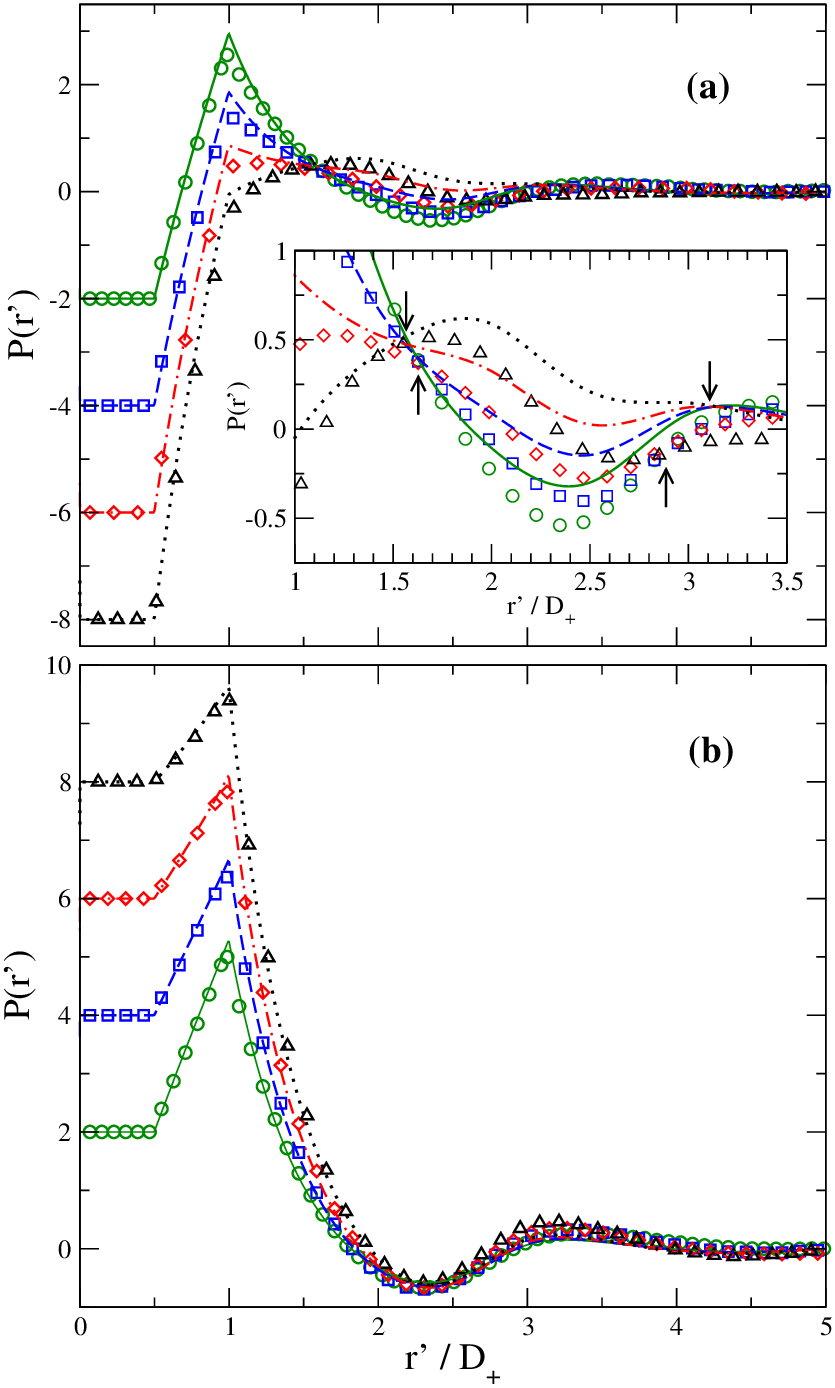}
\caption{Integrated charge of a size-asymmetric 1:1, 1 M electrolyte
  around a charged macroion. 
  In Fig. \ref{fig_z2468_c2_c4}(a) the open circles, squares,
  diamonds and triangles correspond to Monte Carlo simulations for
  $z_M=-2,-4,-6,-8$, respectively, and in Fig. \ref{fig_z2468_c2_c4}(b)
  are associated to $z_M=2,4,6,8$, respectively. In Fig. \ref{fig_z2468_c2_c4}(a) the
  solid, dashed, dot-dashed and dotted lines correspond to HNC/MSA
  results for $z_M=-2,-4,-6,-8$, respectively, and in Fig. 
  \ref{fig_z2468_c2_c4}(b) are associated to $z_M=2,4,6,8$, respectively. 
}
\label{fig_z2468_c2_c4}
\end{center}
\end{figure}

To analyze the behavior of the integrated charge 
as we depart from the point 
of zero charge, in 
Fig. \ref{fig_z2468_c2_c4} 
we present $P(r')$ of MC and HNC/MSA for 
two sequences in $z_M$.
When counterions are smaller than coions ($z_M<0$, 
Fig. \ref{fig_z2468_c2_c4}a), the maximum simulational charge reversal 
observed at the OHP decreases 
as the valence of the macroion becomes more negative. In fact, 
charge reversal within the Helmholtz planes disappear altogether 
for $z_M\approx -7$.
For $z_M < -7$ the first layer of charge reversal 
shifts further away from the macroparticle.
On the other hand, in the curves corresponding to coions smaller than
counterions ($z_M>0$, Fig. \ref{fig_z2468_c2_c4}b) overcharging occurs, and 
the difference between 
the peak of overcharging (at the OHP) and $z_M$ (at the surface) decreases
as $z_M$ increases, which is the result of the growing macroion-coions
repulsion. Overcharging virtually disappears everywhere when this electrostatic repulsion 
becomes strong enough, which for this system corresponds to
$z_M\approx 24$. Consequently, the maximum of overcharging 
only happens at the OHP, indicating that it is a feature directly caused by 
the size asymmetry of the ions in a binary  
electrolyte. From Fig. \ref{fig_z2468_c2_c4} the good coincidence 
between the simulations and the HNC/MSA theory is manifest. 

A more careful inspection of the Monte Carlo IC graphs when $z_M<0$ 
(see the inset of Fig. \ref{fig_z2468_c2_c4}a) shows that, 
at least for short 
distances, the profiles seem to cross each other at specific points, 
defining
in this way intervals where the integrated charge as a function of 
$z_M$ follows a regular pattern, 
as we explain next. Thus, for instance, $0\leq r'/D_+ \lesssim 1.6$ 
(note the upward arrow in the figure) delimits a
region where, for fixed $r'$, the simulated IC 
decreases when $z_M$ diminishes. 
HNC/MSA exhibits the same 
trend for $0 \leq r'/D_+ \lesssim 1.55$ (downward arrow).
The next crossing point of the charge profiles determines another space 
where, for a given position, the IC now {\it augments} when $z_M$ 
decreases. For 
simulations this occurs in the interval $1.6 \lesssim r'/D_+ \lesssim 2.9$ 
(upward arrow), and in $1.55 \lesssim r'/D_+ \lesssim 3.1$ 
for HNC/MSA (downward arrow). Below, we will return to explore the 
relevance of these findings. Previously, we proceed to consider 
the $P(r')$ in URMGC.

\begin{figure}[htbp]
\begin{center}
\includegraphics[width=8.0cm]{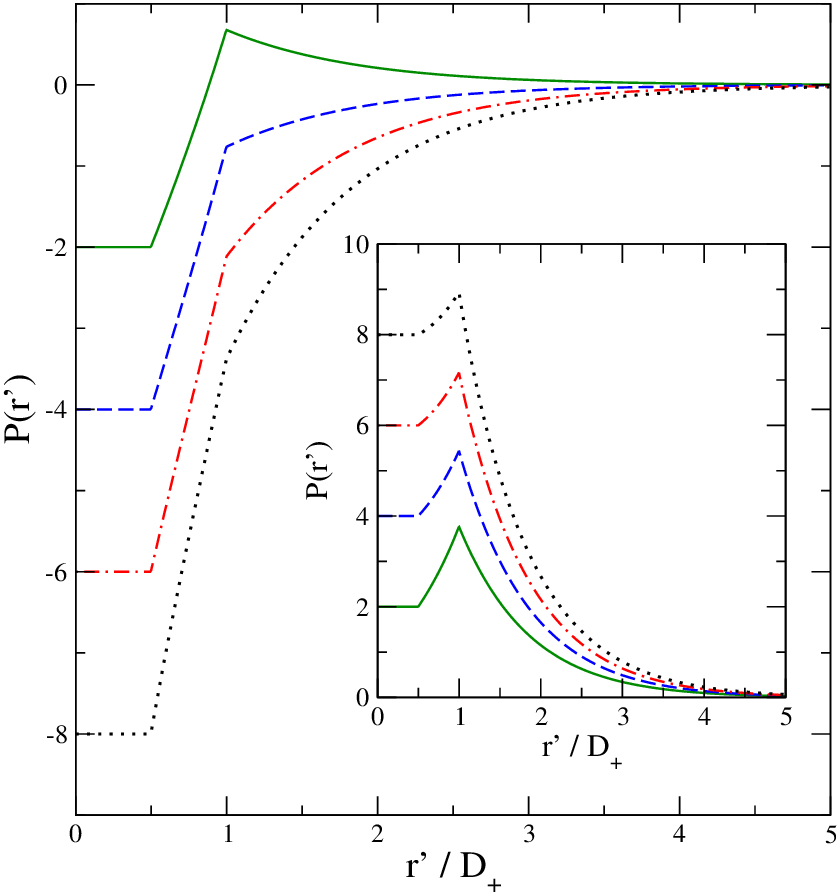}
\caption{The same as in Fig. \ref{fig_z2468_c2_c4} but for the URMGC theory. 
  In the main figure the 
  solid, dashed, dot-dashed and dotted lines correspond to
  $z_M=-2,-4,-6,-8$, respectively, and in the inset to 
  $z_M=2,4,6,8$, respectively.
}
\label{fig_z2468_c2_c4_pb}
\end{center}
\end{figure}

The integrated charge of URMGC does not follow any of the
aforesaid tendencies. In
the first place, for $z_M < 0$ (Fig. \ref{fig_z2468_c2_c4_pb}a), such
theoretical approach exhibits charge reversal only for
a colloidal charge of $z_M = -2$ and this reversal is
weak ($P_{max}(r')\approx 0.7$ at the OHP), which means
that, differently from Monte Carlo and HNC/MSA, the
CR disappears already for $z_M < -2$.
The origin of this comportment of the CR can be traced back to the
very low values of the URMGC concentration
of small cations near a discharged surface reported in the main panel
of Fig. \ref{fig_c4_pm_z0_hs}. When the negative charge 
on the initially neutral colloid starts to
grow (i.e. $z_M$ becomes more negative), the small (and positive) counterions
are attracted to the macroparticle and increase their
number in the Helmholtz zone, then producing charge reversal
for very low $z_M$ ($-2 \lesssim z_M <0$, for URMGC). However, due
precisely to the mentioned URMGC scarcity of small counterions
at $z_M = 0$, the growth
of the negative surface charge overtakes that of the charge reversal
owed to the counterions, and rapidly suppresses it.
Secondly, and in further contrast with MC and HNC/MSA,
in the data of Fig. \ref{fig_z2468_c2_c4_pb}a we notice that, once
the reversal of charge in URMGC ceases to occur in the
Helmholtz region (for $z_M \lesssim -2$), it is never observed again
at any point beyond the OHP in the
monotonically decreasing profiles of $P(r')$.
Evidently, this is another example of the localization
of the ionic size and size asymmetry effects in the semi-punctual URMGC
formalism.
Complementarily, for $z_M > 0$ (Fig. \ref{fig_z2468_c2_c4_pb}b), the URMGC integrated charge
presents overcharging within the Helmholtz planes,
which diminishes slowly as $z_M$ augments.
Notwithstanding, the OC in this theory is less important than that in
simulations and HNC/MSA, as can be verified from a
comparison of the respective differences $|P(r')-z_{M}|$.
Besides, and as expected, the IC curves in 
Fig. \ref{fig_z2468_c2_c4_pb}b go uniformly to zero
after the OHP.
Such abatement of the overcharging in URMGC for
positive surface charges can also be explained
in terms of the low presence of small ions close
to a discharged macroparticle.
In this case, when the positive colloidal charge is being
incremented from the zero value, the shortage of small cations (coions) available
to build-up the overcharging in the proximities
of the surface lessens the magnitude of the anomalous effect.

On the other hand, apart from the usual monotonicity
of the URMGC integrated charge as a function of the distance,
in the two sequences of curves for different $z_M$ presented in Fig. \ref{fig_z2468_c2_c4_pb},
we observe an extra monotonic character of the $P(r')$ profiles,
this time with respect to the variation of $z_M$. Specifically,
in Figs. \ref{fig_z2468_c2_c4_pb}a 
and \ref{fig_z2468_c2_c4_pb}b the IC curves associated to distinct values
of $z_M$ never cross each other. Thus, we find that
the URMGC integrated charge functions, for any fixed $r'$, 
satisfy the condition

\begin{equation}
\frac{\partial P(\sigma_0,r')}{\partial \sigma_0} > 0.
\label{loc}
\end{equation}

\noindent 
Otherwise, and as it was prefigured in relation with the inset of 
Fig. \ref{fig_z2468_c2_c4}a, the integrated charges of MC and HNC/MSA show regions
where the alternative condition 

\begin{equation}
\frac{\partial P(\sigma_0,r')}{\partial \sigma_0} < 0
\label{inloc}
\end{equation}

\noindent is satisfied. We realize that this last ``anomalous" behavior is possible
due to the presence of crossing points in the corresponding IC profiles.
Physically, Eq. 
(\ref{loc}) establishes that if the charge over the macroion 
is increased, the corresponding effect in the spherical EDL is to 
augment locally the IC, which is intuitively awaited. 
Contrastingly, Eq. (\ref{inloc}) states that 
the enlargement of the macroion's charge promotes a local decrease in 
the IC. We refer to this behavior as a 
{\it local inversion of derivative of the integrated charge} (LIDIC).  
It should be noted that such inversion of the derivative occurs 
only when $z_M<0$. This same phenomenon 
seems to be absent for MC and HNC/MSA if $z_M>0$ 
(see Fig. \ref{fig_z2468_c2_c4}b) and, instead, it 
is apparent that all the IC profiles meet at their first minimum 
($r'/D_{+} \approx 2.3$).

\begin{figure}[htbp]
\begin{center}
\includegraphics[width=8.0cm]{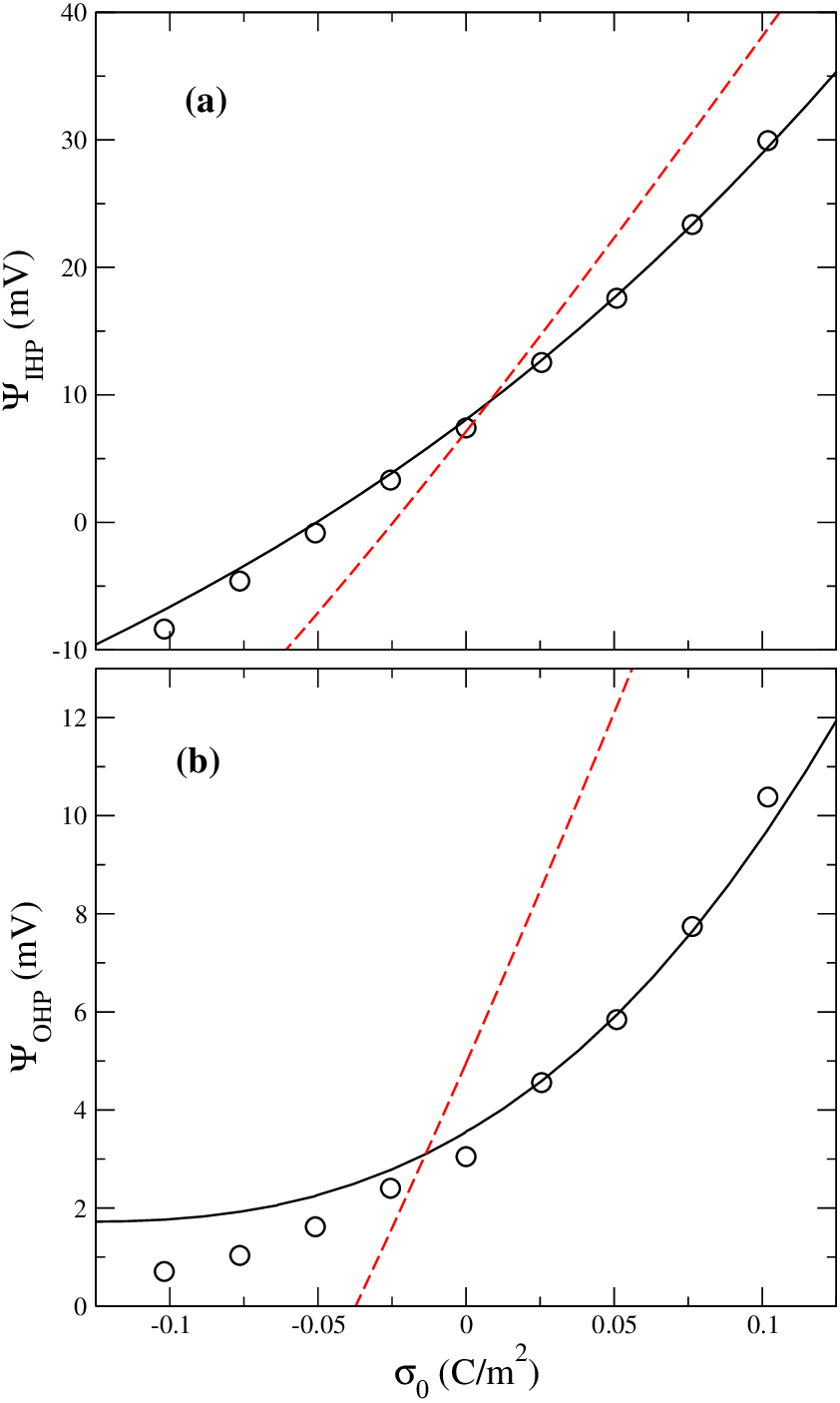}
\caption{Mean electrostatic potential (MEP) at the Helmholtz planes as
  functions of the surface charge density of a macroion immersed in a
  size-asymmetric 1:1, 1 M electrolyte. 
  The open circles, and solid and dashed lines correspond to simulations, and the 
  HNC/MSA and URMGC theories, respectively. 
  In Fig. \ref{zeta_all_c2_c4}(a) the MEP at the IHP is plotted, and in the
  Fig. \ref{zeta_all_c2_c4}(b) the same is done for the MEP at the OHP. 
}
\label{zeta_all_c2_c4}
\end{center}
\end{figure}

In order to explore the behavior of the potential-charge relationship in the 
SEDL, the simulational and theoretical mean electrostatic 
potentials at the Helmholtz planes as functions of $\sigma_0$
are plotted in Fig. \ref{zeta_all_c2_c4}. As it is usual in the
analysis of the $\psi_{IHP}(\sigma_0)$ and $\psi_{OHP}(\sigma_0)$ 
functions, the colloidal 
charge is specified here in terms of the surface density $\sigma_0$ ($=z_M e/(\pi D_M^2)$).
The reader can easily pass from $\sigma_0$ to $z_M$ if realizes
that the symbols (open circles) in the figure correspond to the sequence of
integer valences $z_M=-8,-6,-4,-2,0,+2,+4,+6,+8$.
In Figs.\ref{zeta_all_c2_c4}a and  \ref{zeta_all_c2_c4}b,
we observe that the $\psi_{IHP}(\sigma_0)$ and $\psi_{OHP}(\sigma_0)$
curves for the three approaches, namely Monte Carlo, HNC/MSA and URMGC, display an increasing
monotonic behavior and positive potentials of zero charge (PZCs).
Besides, the simulations and theories predict {\it the existence of intervals
of negative colloidal charges, near the point of zero charge, 
for which the potentials at the IHP and OHP can be positive}, i.e.
where the conditions $\psi_{IHP}\sigma_0 < 0$ and $\psi_{OHP}\sigma_0 < 0$ are accomplished.
As it is indicated by Eq. (\ref{pot2}),
these attributes of  $\psi_{IHP}$ and $\psi_{OHP}$ can be inferred, of course, from the
comportment of
the integrated charge, or better from the function $P(r)/r^2$ (which is
basically the local mean electrostatic field around the macroparticle).
To facilitate our argument
let us start with the case of the URMGC theory.
In Fig. \ref{fig_c4_pm_z0_qr_vr}
we found that, for an uncharged colloid, $P(r') \geq 0$ for all $r'$ in the URMGC scheme. 
Therefore, from the
definition of the MEP (Eq. (\ref{pot2})), it follows that $\psi(r')$ must be
positive at any distance from the neutral surface;
in particular, the relations $\psi_{IHP}>0$ and $\psi_{OHP}>0$ hold. This explains
the positive URMGC PZCs at the IHP and OHP.
Moreover, the series of IC curves graphed in
Fig. \ref{fig_z2468_c2_c4_pb} show a uniform precedence with respect
to the variation of $\sigma_0$, or that $P(r',\sigma_0)>P(r',\sigma'_0)$
if $\sigma_0>\sigma'_0$.
We can state this equivalently by saying that there are not LIDIC regions
in the sequence of curves in Fig. \ref{fig_z2468_c2_c4_pb}.
This necessarily implies that
$\psi(r',\sigma_0)>\psi(r',\sigma'_0)$ at any
distance from the neutral macroparticle,
which elucidates the monotonically increasing behavior
of the $\psi_{IHP}$ and $\psi_{OHP}$ for URMGC noticed in Fig. \ref{zeta_all_c2_c4}.
This same precedence of $P(r',\sigma_0)$ with respect to $\sigma_0$
signifies that, at a fixed distance from the macroparticle,
there must be a $\sigma_0^{crit}$ for which $\psi(r',\sigma_0^{crit})=0$.
Given that in the URMGC formalism such $\sigma_0^{crit}$ happens for both 
the IHP and OHP potentials 
within the range plotted in the figure, we thus find negative values of $\sigma_0$,
close to the point of zero charge, associated to positive values of
$\psi_{IHP}$ and $\psi_{OHP}$.

The properties of the $\psi_{IHP}(\sigma_0)$ and $\psi_{OHP}(\sigma_0)$
obtained from Monte Carlo and HNC/MSA can be analyzed
using similar ideas to those just applied to URMG. The task is more
complex, however, because $P(r)/r^2$ is an alternating
function for both the simulations and HNC/MSA, which would require a more detailed
examination of the positive and negative areas involved
in the evaluation of Eq. (\ref{pot2}).
Fortunately, such process can be clarified if we
note that: (i) in Fig. \ref{fig_c4_pm_z0_qr_vr} the ICs
of MC and HNC/MSA for a neutral colloid have a very high positive peak centered at the OHP,
which contributes more than the adjacent minimum at $r'\approx 2.4$ to the
integral of $P(r)/r^2$ when $\psi_{IHP}$ and $\psi_{OHP}$ are being calculated,
and (ii) that precisely these first maximum and minimum
in $P(r')$ dominate the
integrals associated to
$\psi_{IHP}$ and $\psi_{OHP}$ ($\psi_{IHP}$-integral
and $\psi_{OHP}$-integral, respectively).
From (i) and (ii) the positive sign of
the Monte Carlo and HNC/MSA PZCs then arises.
The recognition of the determinant role that the first extrema in the
ICs have for the evaluation of
$\psi_{IHP}$ and $\psi_{OHP}$
is very important since it provides the key to understand the complete 
shape of the $\psi_{IHP}(\sigma_0)$ and $\psi_{OHP}(\sigma_0)$
coming from MC and HNC/MSA.
The rational proceeds as follows.
Differently from URMGC, in Figs. \ref{fig_z2468_c2_c4}a and \ref{fig_z2468_c2_c4}b  
the sequences of ICs
of simulation and HNC/MSA for
distinct colloidal charges
are non-monotonic with respect to $r'$ and display LIDIC regions (Fig. \ref{fig_z2468_c2_c4}a). 
Notwithstanding, in Fig. \ref{fig_z2468_c2_c4}b it can be seen that, for $\sigma_0 >0$,
the height of the overcharging peak
at the OHP increases when $\sigma_0$ augments, or else that the
fact (i) of the previous paragraph occurs in a more
emphatic way as $\sigma_0$ heightens. Consequently, such enhancement of
(i), combined with the prevailing fact (ii), results in the
increasing behavior of $\psi_{IHP}(\sigma_0)$ for
positive $\sigma_0$ observed in simulations and HNC/MSA.
Otherwise, for $\sigma_0 < 0$, it is better to consider 
the OHP as a reference point to split the 
contributions of $P(r')$ to the $\psi_{IHP}$-integral.
In this form, it can be seen that,
for $\sigma_0 \lesssim -0.051$ C/m$^2$ ($z_M \lesssim -4$),
the area comprised by that section of the IC
before the OHP becomes more negative,
and more important than the area
enclosed by the IC beyond the OHP, when
$\sigma_0$ decreases. Thus, such argument explicates
the increasing comportment of $\psi_{IHP}(\sigma_0)$
for negative $\sigma_0$, as well as the existence in Monte Carlo 
and HNC/MSA of
an interval of negative surface charges for which
$\psi_{IHP}$ is positive.
With respect to $\psi_{OHP}(\sigma_0)$ in simulations and HNC/MSA, 
from Fig. \ref{fig_z2468_c2_c4} it is evident that, for those 
1:1 systems, the value (and sign !) 
of the $\psi_{OHP}$-integral is determined by 
overcharging, for $\sigma_0 >0$, and by 
charge reversal, for $\sigma_0 <0$. 
For positive colloidal charges, the increment of the 
height of the overcharging peak at OHP when $z_M$ augments 
dictates the increasing positiveness 
of $\psi_{OHP}$. In addition, we notice that, for negative 
macroparticle charges, a dominant peak of charge reversal occurs either at the OHP 
or beyond that point, which in the present work is a distinctive feature of 
those descriptions that incorporate consistently the ionic size and size asymmetry 
effects (i.e. simulations and HNC/MSA) and, therefore, it is absent in URMGC. 
Such strong charge reversal 
gives rise to the positive sign of $\psi_{OHP}$ for negatively charged colloids, and 
then completes our explanation
of the fact that, for simulations and HNC/MSA, 
$\psi_{OHP}$
is always positive for the range covered in
Fig. \ref{zeta_all_c2_c4}b.

In connection with electrophoresis experiments, and depending on 
the location of the shear plain, 
our results suggest two possible scenarios. 
If the sign of the electrophoretic mobility, $\mu$, were associated   
to the sign of the potential 
$\psi_{IHP}$, our treatment of the size-asymmetric EDL
would predict the inversion of the colloidal mobility 
very near to the point of zero charge. In contrast, if the sign of 
$\mu$ were associated to the sign of the potential $\psi_{OHP}$, our study 
would foresee positive mobilities in 
a wider range of colloidal charges around zero. 
In addition, Fig. \ref{zeta_all_c2_c4}b 
resembles a situation reported in Fig. 2b of Ref. \cite{Jo01} 
by Johnson et al., where zeta potential measurements of 
$\alpha$-alumina in presence of 1 M LiNO$_3$ are plotted as a function 
of pH. In those results, the zeta potential is always positive 
in a wide interval of pH that encompasses the point of zero charge, 
which parallels 
our results in  Fig. \ref{zeta_all_c2_c4}b. This would indicate
that such positive values of the zeta potential are due in part
by the  preferential adsorption of one of the species (Li$^+$),
which in turn should be induced by the ionic size asymmetry,
as deduced from our survey of the spherical EDL. Although other 
complex mechanisms
are into play in these kind of experiments \cite{Jo02,Co01,Gie01,Be01}, 
our size-asymmetric model
seems to capture adequately relevant phenomena occurring in 
real systems, and hence it could
be considered as a basic representation of the EDL to which further 
improvements (as Van der Waals dispersion forces or more
sophisticated chemical mechanisms) can be incorporated in order to predict
experimental data more accurately.

\subsection{Divalent size-asymmetric electrolytes}

We proceed to investigate the properties of the electrical double layer 
for a macroion
immersed in a 2:2 salt, i.e. for systems 
with stronger charge correlations.
As it will be shown, many of the findings reviewed in the prior section 
devoted to univalent electrolytes 
are also present, in an enhanced way, for the case of 
divalent ions. As before, 
we examine first the
EDL at the point of 
zero charge and later we ponder instances with charged macroparticles.
In what follows, we consider EDL systems formed by a colloid and a bath of a  
2:2, 0.5 M electrolyte, with the same diameters as specified in the earlier section.

\begin{figure}[htbp]
\begin{center}
\includegraphics[width=8.0cm]{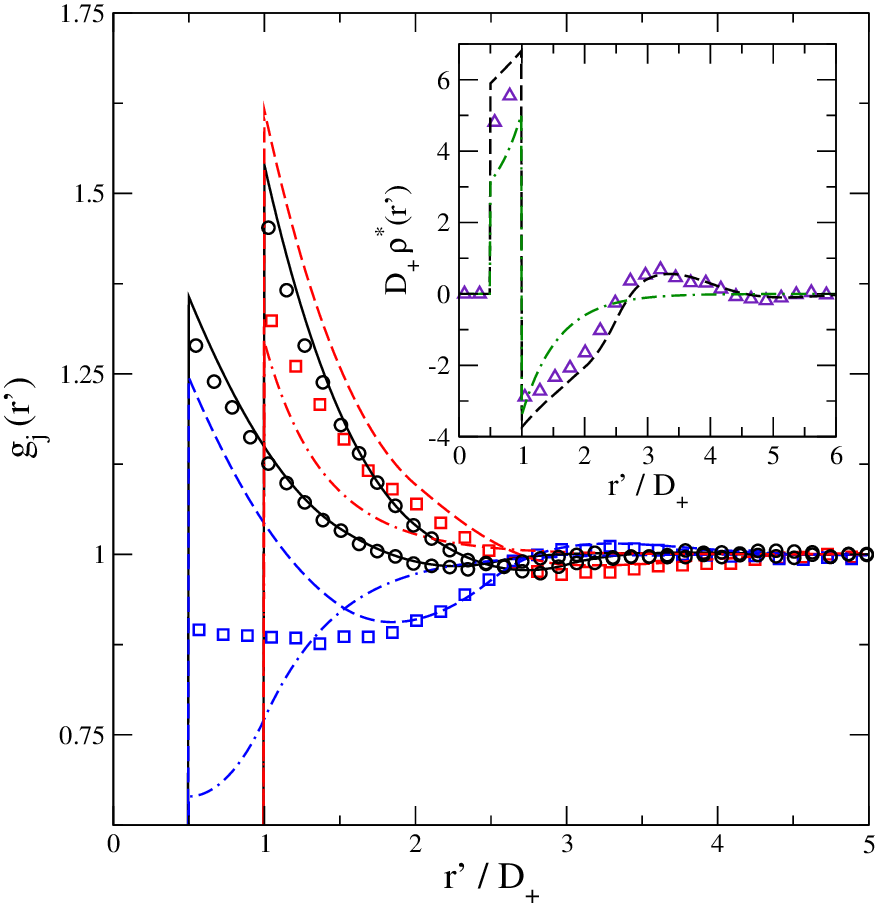}
\caption{Radial distribution functions of a size-asymmetric binary mixture of
  hard spheres and a size-asymmetric 2:2 electrolyte, both at a 0.5 M
  concentration, around an uncharged ($z_M=0$) macroparticle (in the main panel). 
  The small hard spheres 
  have the same diameter as that of the cations, $D_{1}=D_{+}$, and the large hard spheres 
  diameter is the same as that of the anions, $D_{2}=D_{-}$. 
  The open circles 
  correspond to Monte Carlo (MC) simulations of the size-asymmetric 
  mixture of hard spheres and the open squares represent the MC data 
  of the size-asymmetric divalent electrolyte. The
  solid lines are associated to the asymmetric mixture of hard spheres in
  the HNC/MSA approach. The dashed and dot-dashed lines 
  are the HNC/MSA and URMGC results for the 2:2 size-asymmetric 
  salt, respectively. In the inset we depict the corresponding local electrolyte 
  charge density for the 2:2 electrolyte case. Open triangles, 
  and dashed and dot-dashed lines correspond to simulations, and the HNC/MSA 
  and URMGC theories, respectively. 
}
\label{fig_c8_pm_z0_hs}
\end{center}
\end{figure}

The simulational, HNC/MSA and URMGC radial distributions 
of divalent ions around a macroparticle with 
$z_M=0$ are included in Fig. \ref{fig_c8_pm_z0_hs} (main panel). There 
we have incorporated the Monte Carlo 
and HNC/MSA pair correlation functions 
for the associated 
hard-sphere mixture (uncharged ``ions"). 
Again, the idea is to exemplify how the coulombic 
correlations modify the structure of the pure hard-sphere fluid, in order to  
gain some insight into the relative importance of the entropic and charge 
correlations. From the direct contrast between the simulational RDFs of 
hard spheres and ions in the main panel of the figure, 
we observe that the charge effects are 
very strong, 
completely modifying the accumulation of the smaller species around the 
macroparticle. In fact, the changes in the Monte Carlo $g_j(r')$ 
of divalent ions with respect to 
the RDFs of hard spheres are bigger than 
those occurring in univalent systems (compare the main panels of 
Figs. \ref{fig_c4_pm_z0_hs} and \ref{fig_c8_pm_z0_hs}).  
For the simulations of 2:2 electrolytes, the impact of the valence 
is much more important for the smaller ionic species, where
we observe that the corresponding contact population depletes so much that the 
concentration of cations near the surface is below the bulk value. 
A similar dewetting has been reported in a recent study of the 
planar EDL for size-asymmetric
1:1, 1:2 and 1:3 salts \cite{Gi01}, with the multivalent ions 
corresponding to the smaller species. 
In that work it was found that the 
amount of small cations in the proximities of a neutral plane decreased as long as the 
electrostatic coupling (i.e. the valence) was heightened, 
being our results consistent with 
such behavior. 
On the other hand, in Fig. \ref{fig_c8_pm_z0_hs} (main panel) 
we can appreciate that the HNC/MSA results for hard spheres  
compare well with the simulations, whereas for the divalent case 
this formalism overestimates the Monte Carlo data for the two ionic species. 
Furthermore, HNC/MSA fails to describe the correct
tendency for the RDF of big anions, predicting an increase of such 
function with respect to the corresponding RDF of the larger hard spheres, 
which is clearly not the behavior seen 
in the MC data. Despite this poor achievement in the estimation of the $g_j(r')$ 
for the electrolyte, HNC/MSA does well at the level of the local 
electrolyte charge density (see inset of Fig. \ref{fig_c8_pm_z0_hs}), capturing the correct
trends of the simulations. 
This is not a surprising circumstance given that the dependence of $\rho^*(r')$ 
on the difference $g_{+}(r')-g_{-}(r')$ explains 
why 
the notable discrepancies existing 
between the RDFs of Monte Carlo and 
HNC/MSA are attenuated when the local charge densities are compared instead.
URMGC, on the other hand, highly disagrees with the simulations 
for the $g_j(r')$ and $\rho^*(r')$. Special note must be taken 
of the very low URMGC concentration 
of small ions in the Helmholtz zone, which, analogously to the 
univalent case, will determine the weak intensity of the CR and 
OC phenomena for 2:2 systems in this theory.

\begin{figure}[htbp]
\begin{center}
\includegraphics[width=8.0cm]{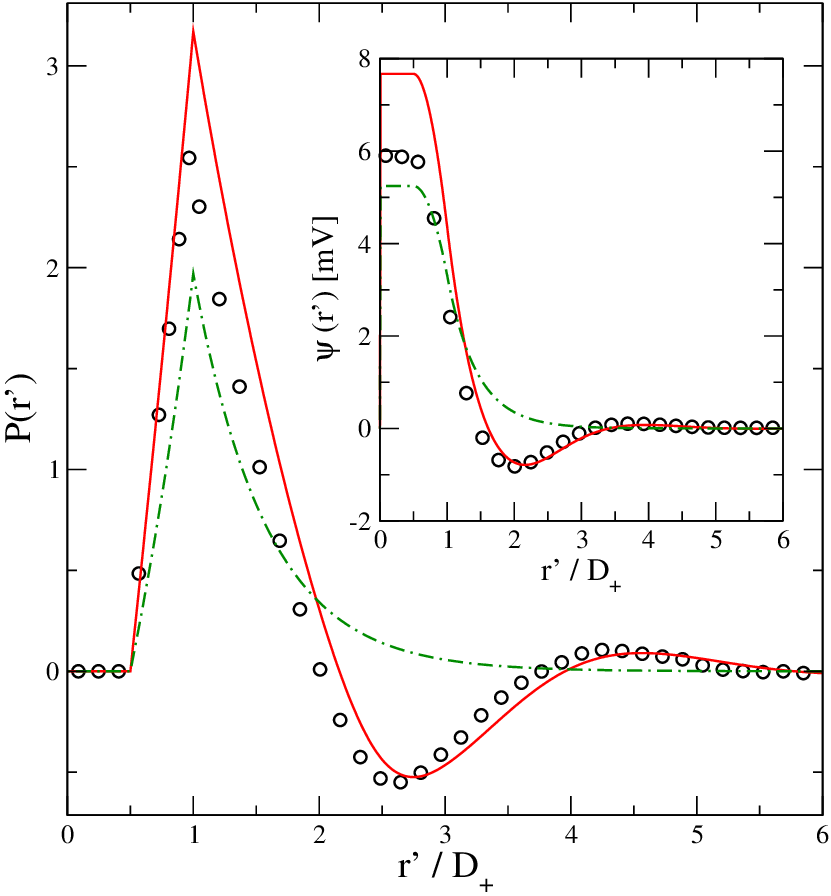}
\caption{Integrated charge (main panel) and mean electrostatic potential (inset) of a 
  size-asymmetric 2:2, 0.5 M electrolyte around an uncharged ($z_M=0$)
  macroparticle. 
  The open circles, solid and
  dot-dashed lines are associated to the MC, HNC/MSA and URMGC data,
  respectively.  
}
\label{fig_c8_pm_z0_qr_vr}
\end{center}
\end{figure}

The simulational and theoretical ionic distributions of the 
divalent salt yield the accumulated charge
and mean electrostatic potential given in Fig. 
\ref{fig_c8_pm_z0_qr_vr} and its inset. As it was 
detected in the monovalent situation, the ionic size asymmetry promotes an adsorbed 
layer of charge and the concomitant existence of a zero 
surface charge double layer (see Fig. \ref{fig_c8_pm_z0_hs}), as well as a non-zero MEP 
at (and between) the Helmholtz planes.
The adsorbed charges, up to the OHP, are $\approx 2.6e$, $3.3e$, and $2.0e$ 
for MC, HNC/MSA, and URMGC, respectively, which are 
smaller than those observed for the 1:1 salt. 
In turn, contrasted with Monte Carlo, 
the MEP within the Helmholtz planes is overestimated 
by HNC/MSA and underestimated by URMGC. 
Globally we see that the performance of HNC/MSA and URMGC 
is very similar to that displayed for the monovalent salt, with a better 
qualitative similitude between HNC/MSA and the simulations. 
For URMGC the $P(r')$ and $\psi(r')$ profiles evidence 
the restriction of all the size asymmetry and hard-core effects to happen uniquely 
in the Helmholtz zone.

\begin{figure}[htbp]
\begin{center}
\includegraphics[width=8.0cm]{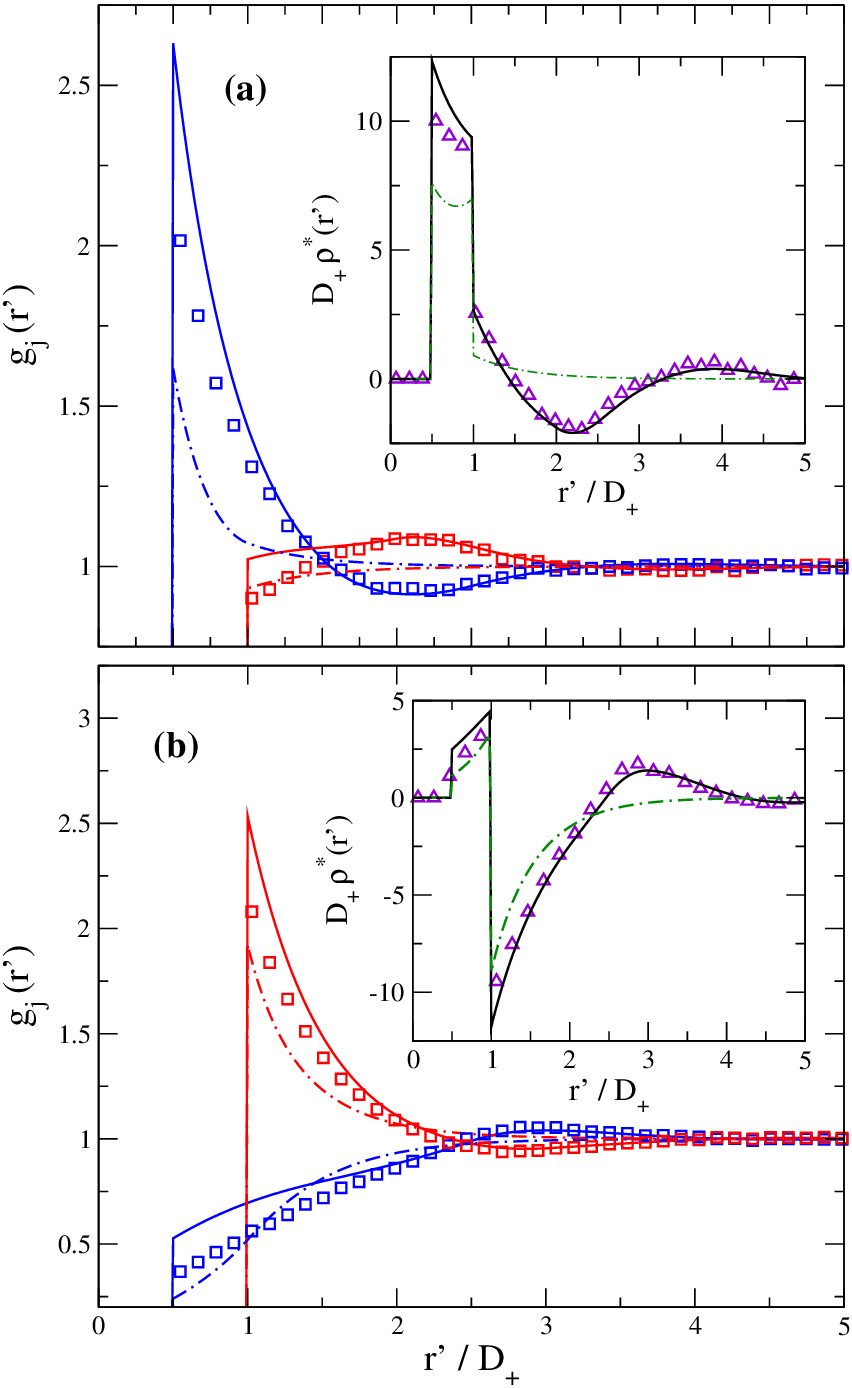}
\caption{Radial distribution functions of a size-asymmetric 2:2, 0.5 M
  electrolyte,  around a charged macroion (main panel). 
  In Fig. \ref{fig_z4_c6_c8}(a) $z_M=-4$, and in Fig. \ref{fig_z4_c6_c8}(b) $z_M=4$. 
  The open squares, solid and dot-dashed lines represent the MC, 
  HNC/MSA and URMGC results, respectively. In the insets the corresponding local electrolyte 
  charge densities are portrayed. Open triangles, 
  and solid and dot-dashed lines are associated to simulations, and the HNC/MSA 
  and URMGC theories, respectively. 
}
\label{fig_z4_c6_c8}
\end{center}
\end{figure}

\begin{figure}[htbp]
\begin{center}
\includegraphics[width=8.0cm]{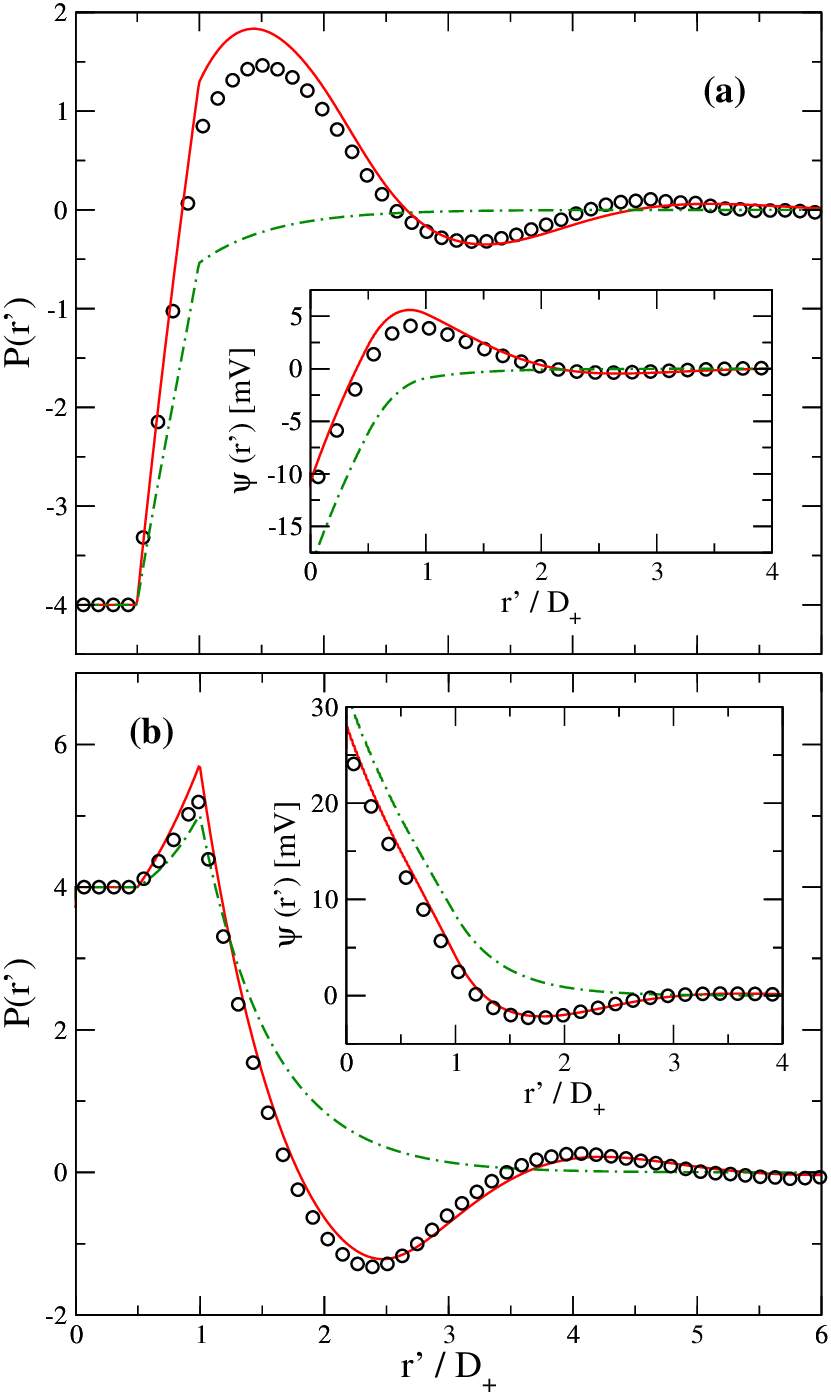}
\caption{Integrated charge (main panel) and mean electrostatic potential (inset) of a 
  size-asymmetric 2:2, 0.5 M electrolyte around a charged 
  macroion. 
  In Fig. \ref{fig_z4_c6_c8_qr_vr}(a) $z_M=-4$, and in Fig. \ref{fig_z4_c6_c8_qr_vr}(b)
  $z_M=4$.
  The open circles, solid and
  dot-dashed lines correspond to the MC, HNC/MSA and URMGC data,
  respectively.   
}
\label{fig_z4_c6_c8_qr_vr}
\end{center}
\end{figure}

The Monte Carlo and theoretical RDFs, LECD, IC, and MEP curves 
corresponding to the divalent system 
and two values of the colloidal charge, namely $z_M=-4$ and $z_M=4$, 
are plotted in 
Figs. \ref{fig_z4_c6_c8}  and \ref{fig_z4_c6_c8_qr_vr}.
When $z_M=-4$ (Fig. \ref{fig_z4_c6_c8}a), 
there is a strong 
adsorption of counterions inside the Helmholtz planes, 
accompanied by an important depletion of coions, particularly for 
MC and HNC/MSA (main panel). 
Besides, it is worth noticing that very close to the OHP the concentration 
of coions is smaller than that of counterions (compare with Fig. \ref{fig_z4_c2_c4}a),
which seems to indicate that at these
distances the repulsion forces between the macroion and the coions are far 
from being screened. 
The associated local charge densities are portrayed in the inset, 
where we have curves that do not jump to a negative value immediately after 
the OHP, showing that the macroion-coions repulsion 
propagates beyond the Helmholtz planes. 
When $z_M=4$ (Fig. \ref{fig_z4_c6_c8}b), 
and despite of the intense electrostatic repulsion,
the MC, HNC/MSA and URMGC concentrations of the small coions close to the IHP 
turn out to be small but finite, effectively increasing the charge of 
the macroion, and therefore driving a significative enhancement in the counterions 
concentration near the OHP (main panel). As a result, the LECD profiles develop
a small first positive layer followed by an ample negative region 
(see inset).

In Fig. \ref{fig_z4_c6_c8_qr_vr} we present the integrated charges and 
MEPs for the divalent salt and macroion valences $z_M = -4$ and $z_M=4$. 
For $z_M=-4$ (panel (a)), 
the ICs of Monte Carlo and HNC/MSA exhibit charge reversal in the Helmholtz 
zone, but the 
maximum charge reversals are outside that region (at $r'/D_+ \approx 1.5$ and $r'/D_+ \approx 1.4$ 
for simulations and HNC/MSA, respectively). 
The corresponding MEPs, on the other hand, display a strong reversal,
with the maximum potential being inside the Helmholtz planes (see inset). In the 
case $z_M=4$ (panel (b)) 
we see that the MC and HNC/MSA integrated charge profiles present overcharging, 
followed by alternate 
oscillations of decreasing amplitude, whereas the associated MEPs (in the inset) decay 
in a fluctuating manner to zero. This observation of OC in the computer ``experiments" of 
divalent systems complements our previous findings for univalent salts and, thus,  
consolidates our simulational proof of overcharging as a genuine feature of 
size-asymmetric primitive model EDLs. 

A simultaneous analysis of all the structural information for 
divalent systems contained in 
Figs. \ref{fig_z4_c6_c8}  and \ref{fig_z4_c6_c8_qr_vr}
show that the HNC/MSA theory 
follows closely the Monte Carlo data, whereas 
the URMGC approach exhibits notable differences with respect to the 
simulations. 
From these figures it is 
also verified that in the URMGC description of the EDL there is 
a confinement of the steric-related phenomenology (e.g. 
CR, OC and the non-monotonic character of the structural functions) 
within the Helmholtz zone.

\begin{figure}[htbp]
\begin{center}
\includegraphics[width=8.0cm]{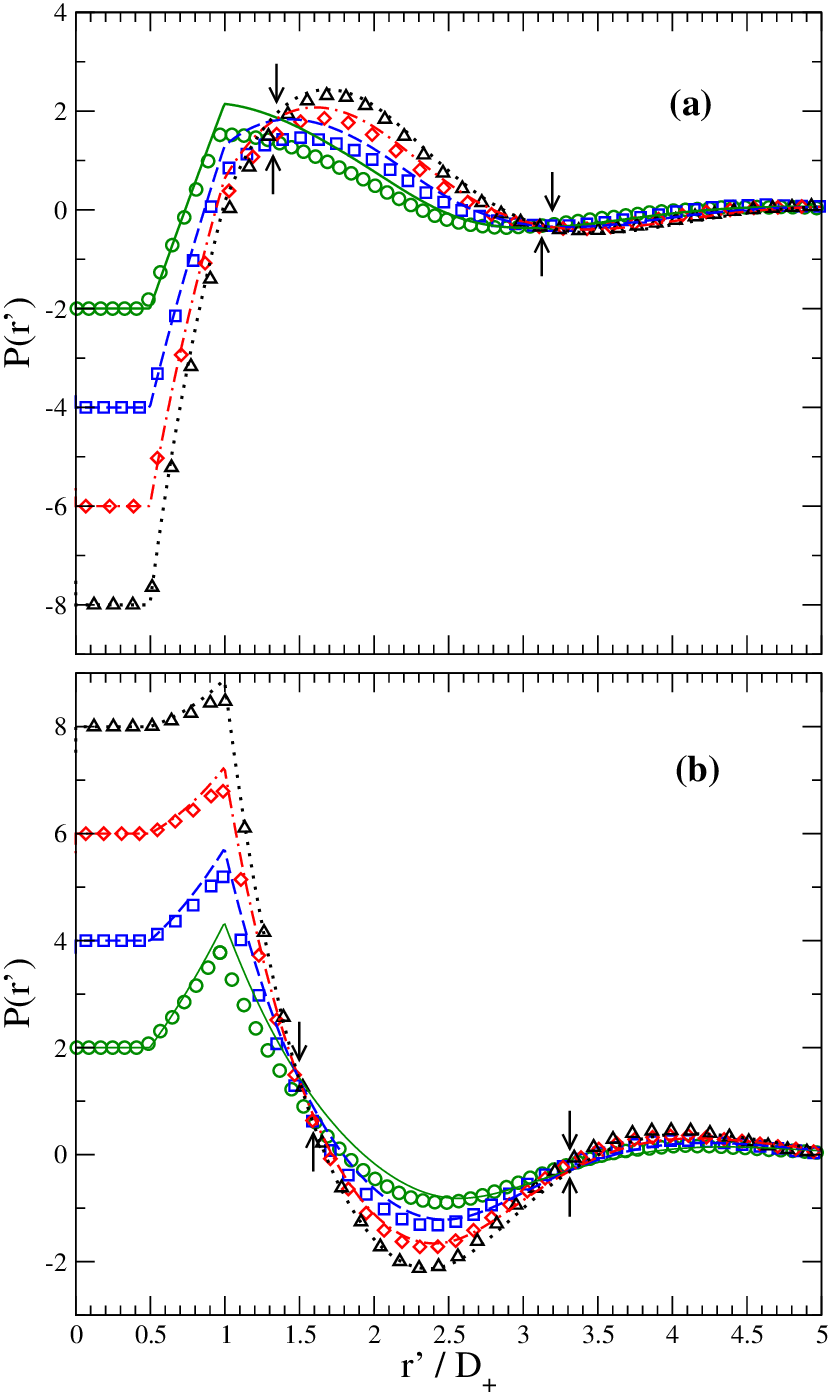}
\caption{Integrated charge of a size-asymmetric 2:2, 0.5 M electrolyte
  around a charged macroion. 
  In Fig. \ref{fig_z2468_c6_c8}(a) the open circles, squares,
  diamonds and triangles correspond to Monte Carlo simulations for
  $z_M=-2,-4,-6,-8$, respectively, and in Fig. \ref{fig_z2468_c6_c8}(b)
  are associated to $z_M=2,4,6,8$, respectively. In Fig. 
  \ref{fig_z2468_c6_c8}(a) the 
  solid, dashed, dot-dashed and dotted lines correspond to HNC/MSA
  results for $z_M=-2,-4,-6,-8$, respectively, and in Fig. 
  \ref{fig_z2468_c6_c8}(b) are associated to $z_M=2,4,6,8$, respectively. 
}
\label{fig_z2468_c6_c8}
\end{center}
\end{figure}

Fig. \ref{fig_z2468_c6_c8} portrays the Monte Carlo and HNC/MSA integrated charges 
for varying $z_M$ ($-8 \leq z_M \leq 8$) in order 
to analyze the evolution of the size asymmetry effects when the 
divalent systems depart from 
the point of zero colloidal charge. For $z_M< -2$ (Fig. \ref{fig_z2468_c6_c8}a) 
the IC profiles of MC and HNC/MSA present charge reversal, 
with the maximum located outside the Helmholtz zone. As the charge increases 
(towards $z_M=0$) the location of the peak of each curve shifts to shorter distances
and eventually lands at the OHP. Further increase of $z_M$ 
leads to the results plotted in Fig. \ref{fig_z2468_c6_c8}b, where the
system displays overcharging, with a maximum intensity (i.e. $|P(r')-z_M|$) that decreases with the augment of the 
macroion's valence.
This overcharging phenomenon only takes place in between the Helmholtz planes, and it is followed by a 
first region of charge reversal. At the limiting valence $z_M\approx 16$
overcharging virtually disappears; meanwhile the charge reversal beyond the OHP
continues to exist with an increasing magnitude. 
All these characteristics of $P(r')$ for $z_M >0$ 
are equally described for both the simulations and HNC/MSA. 
The complete development of the $P(r')$ for 2:2 electrolytes 
and $-8 \leq z_M \leq 8$ reveals a passage from charge reversal to overcharging.  
This crossover at $z_M = 0$ is due uniquely to the 
size asymmetry of the salt ions, as the change of sign of 
$z_M$ merely inverts the role of the anions and cations (as coions or counterions). 

Therefore, in the present investigation it is has been found that the charge reversal and overcharging observed 
near the point of zero charge are mainly caused by 
entropic contributions coming from 
the size-asymmetric nature of the electrolyte ions. However, 
when $z_M$ 
is increased, the charge reversal and overcharging in the Helmholtz zone 
disappear, and only CR persists. 
This remaining charge reversal beyond the OHP comes from the interplay between excluded volume and 
electrostatic correlations  
and, thus, it is not exclusive of a size-asymmetric model. 
That is the reason why 
charge reversal has been already reported 
in many studies of the restricted primitive model EDL, in which 
the excluded volume effects 
are also consistently taken into account \cite{Te01}. 
Accordingly, EDL theories for genuine punctual ions and uniformly charged 
surfaces do not predict charge reversal at all.

We would like to mention that our divalent systems
also exhibit regions of local inversion of the derivatives of the integrated charge  
(LIDIC). These  
zones are delimited in Figs. \ref{fig_z2468_c6_c8}a and 
\ref{fig_z2468_c6_c8}b by the upward arrows for Monte Carlo and by the downward 
arrows for HNC/MSA. Inside these 
``anomalous" regions 
the accumulated charge at any point decreases when the 
colloidal charge is augmented and vice versa. 
Similarly to the univalent case, and starting from the definition 
of the MEP (see Eq. (\ref{pot2})), 
the existence of this LIDIC domains provides a way to explain 
the behavior of the potential-charge relationship at the Helmholtz 
planes.

\begin{figure}[htbp]
\begin{center}
\includegraphics[width=8.0cm]{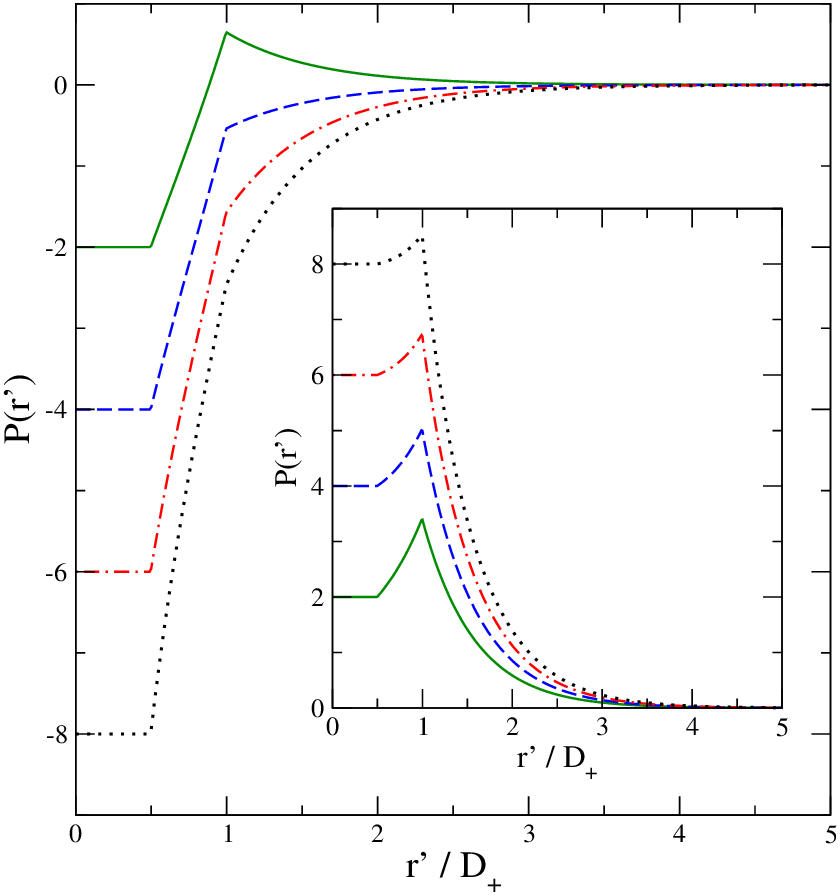}
\caption{The same as in Fig. \ref{fig_z2468_c6_c8} but for the URMGC theory. 
  In the main figure the  
  solid, dashed, dot-dashed and dotted lines correspond to
  $z_M=-2,-4,-6,-8$, respectively, and in the inset to 
  $z_M=2,4,6,8$, respectively.
}
\label{fig_z2468_c6_c8_pb}
\end{center}
\end{figure}

The corresponding sequence ($-8 \leq z_M \leq 8$) of URMGC integrated charges for 2:2 
electrolytes is given in Fig. \ref{fig_z2468_c6_c8_pb}. Visibly, each one of the profiles 
in this series variates monotonically with respect to the distance $r'$ and, as a whole, the 
full set of curves changes monotonically as a function of $z_M$. In other words, after 
the OHP, every $P(r')$ goes to zero without spatial oscillations, 
and the complete sequence does not show LIDIC regions. 

\begin{figure}[htbp]
\begin{center}
\includegraphics[width=8.0cm]{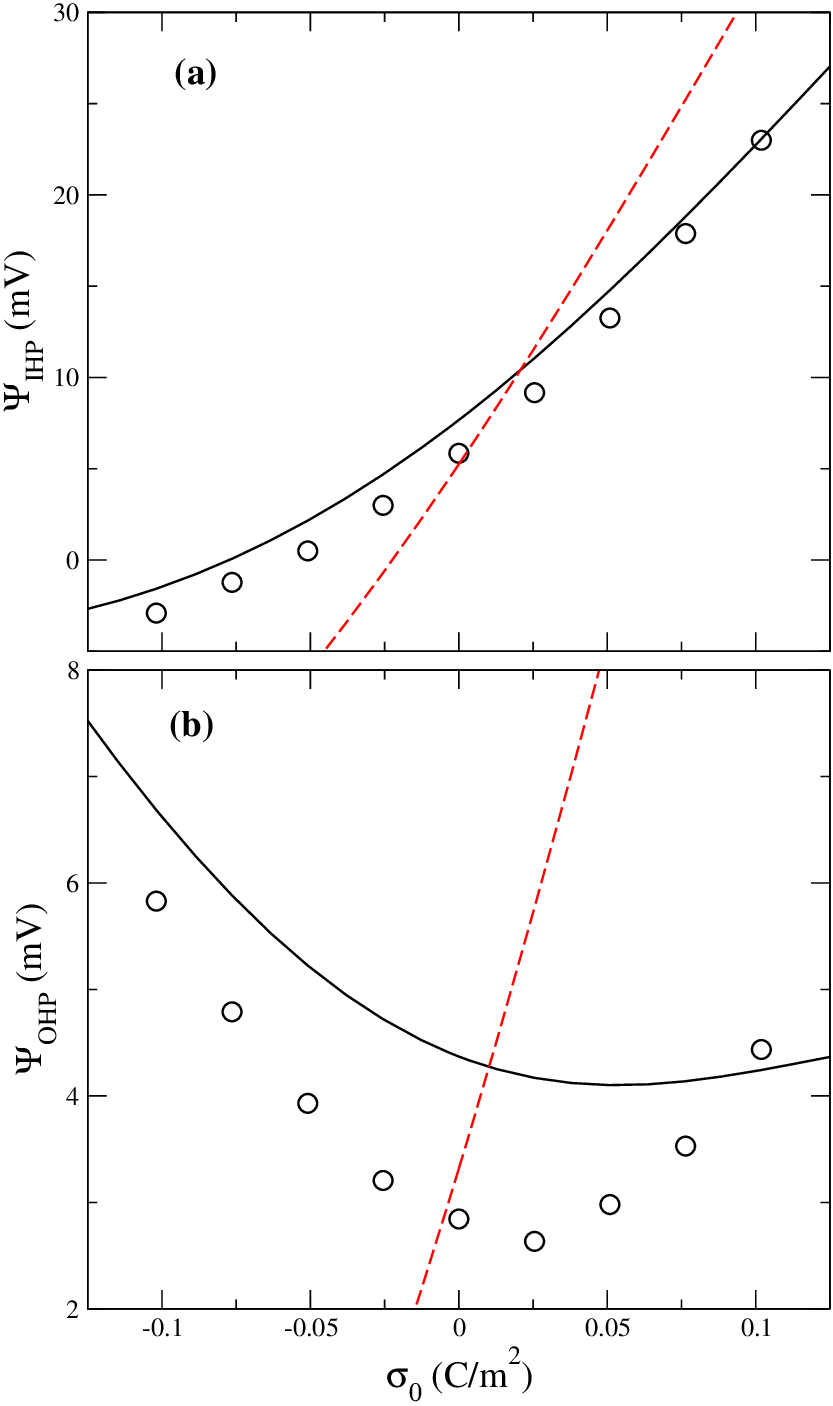}
\caption{Mean electrostatic potential (MEP) at the Helmholtz planes as
  functions of the surface charge density of a macroion immersed in a
  size-asymmetric 2:2, 0.5 M electrolyte. 
  The open circles, and solid and dashed lines correspond to simulations, and the 
  HNC/MSA and URMGC theories, respectively.
  In Fig. \ref{zeta_all_c6_c8}(a) the MEP at the IHP is plotted, and in the
  Fig. \ref{zeta_all_c6_c8}(b) the same is done for the MEP at the OHP. 
}
\label{zeta_all_c6_c8}
\end{center}
\end{figure}

The MEP of MC, HNC/MSA, and URMGC at the Helmholtz planes
are shown in Fig. \ref{zeta_all_c6_c8} as a function of the surface
charge $\sigma_0$. At the IHP, all the approaches exhibit a MEP that
is monotonic and displays a positive zero--charge potential and a negative
$\sigma_0^{crit}$ such that $\psi_{IHP}>0$ for
$\sigma_0^{crit}<\sigma_0<0$, just like in the monovalent systems.
From panel (b) we also find that  $\psi_{OHP}$ is also monotonic
for URMGC. In contrast to the monovalent case,
however, $\psi_{OHP}$ from MC and HNC/MSA is non-monotonic and displays a
minimum at a positive $\sigma_0^{min}$. This augment of $\psi_{OHP}$
when $\sigma_0$ decreases below $\sigma_0^{min}$ suggests that there could be
experimental systems in which the reverted electrophoretic mobility
always increases its magnitude.

The trends of these MEPs can be explained in terms of the corresponding
IC profiles (see Figs. \ref{fig_z2468_c6_c8} and \ref{fig_z2468_c6_c8_pb}), 
as discussed previously in the context of monovalent 
electrolytes. In particular, the concavity of $\psi_{OHP}$ for MC and HNC/MSA 
can be explained by observing in Fig. \ref{fig_z2468_c6_c8}b 
the behavior of the integral of 
$P(r)/r^2$ between two consecutive electroneutral points of $P(r)$,
and noting
that the absolute value of these negative areas decreases when $z_M$ 
diminishes.   
Thus, for $\sigma_0 \geq 0$, the area from the OHP to the first
electroneutral point is positive and larger than the negative area
between the first and second electroneutral points
(see Fig. \ref{fig_z2468_c6_c8}b).
Since the difference between these two contributions to the $\psi_{OHP}$
increases with $\sigma_0$, this explains the positive value of
$\psi_{OHP}$ and its increasing behavior. An analogous argument
explains the trends observed when $\sigma_0<0$  (Fig. \ref{fig_z2468_c6_c8}a),  
despite the disappearance 
of the maximum charge reversal at the OHP when $z_M$ decreases.

\begin{figure}[htbp]
\begin{center}
\includegraphics[width=8.0cm]{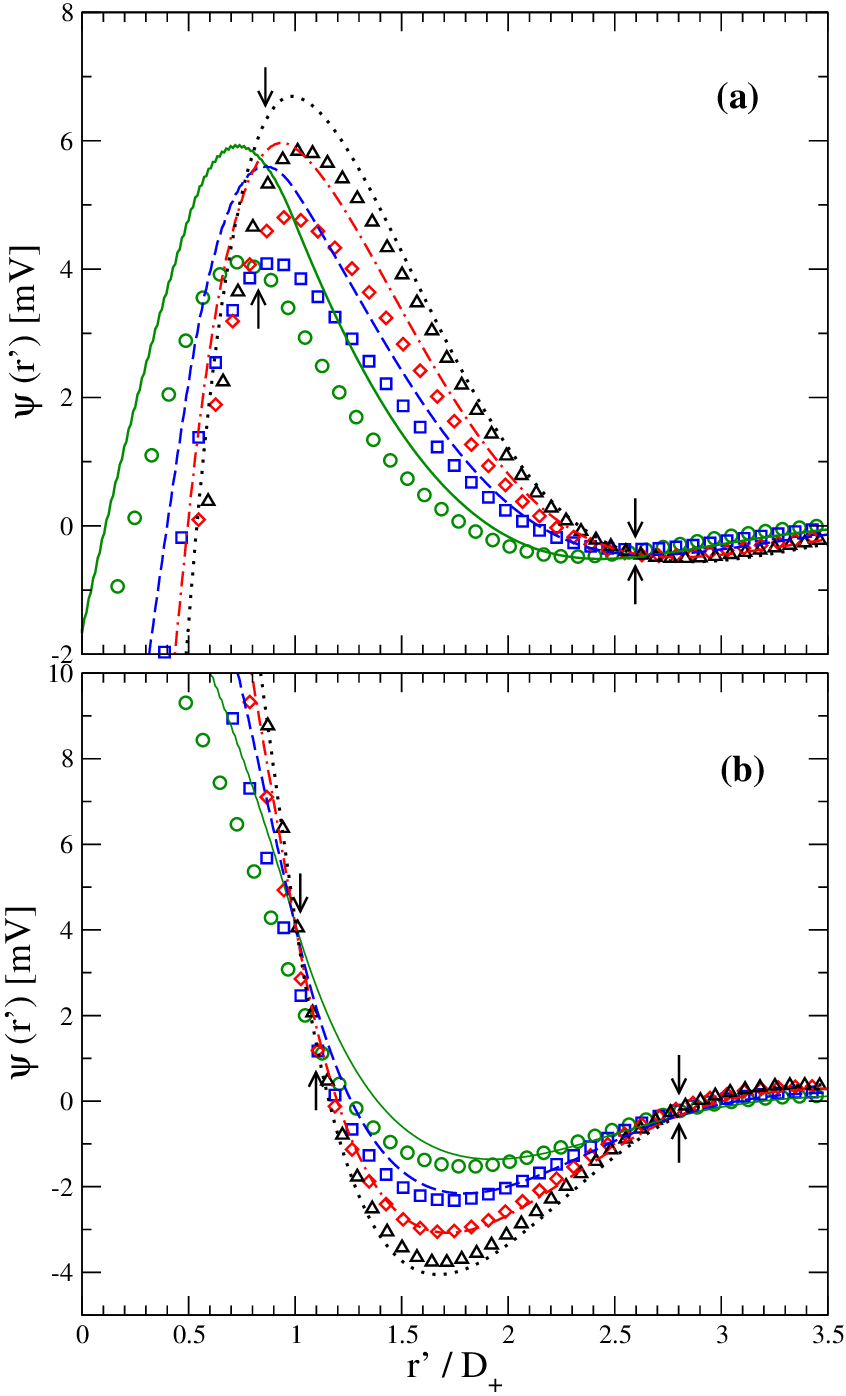}
\caption{Mean electrostatic potential of a size-asymmetric 2:2, 0.5 M
  electrolyte around a charged macroion. 
  In Fig. \ref{fig_v_z2468_c6_c8}(a) the open circles, squares,
  diamonds and triangles correspond to Monte Carlo simulations for
  $z_M=-2,-4,-6,-8$, respectively, and in Fig. \ref{fig_v_z2468_c6_c8}(b) 
  are associated to $z_M=2,4,6,8$, respectively. In Fig. 
  \ref{fig_v_z2468_c6_c8}(a) the 
  solid, dashed, dot-dashed and dotted lines correspond to HNC/MSA
  results for $z_M=-2,-4,-6,-8$, respectively, and in Fig. 
  \ref{fig_v_z2468_c6_c8}(b) are associated to $z_M=2,4,6,8$, respectively. 
}
\label{fig_v_z2468_c6_c8}
\end{center}
\end{figure}

The concavity of $\psi_{OHP}$ can also be cast in terms
of the {\it local inversion of derivative of the MEP} (LIDMEP),
which we  define analogously
to the LIDIC. With this in mind, Fig. \ref{fig_v_z2468_c6_c8}
shows the MEP of the divalent system for several values of
$z_M$. For $z_M<0$ the MC data shows a region of LIDMEP in
$0.82\lesssim r'/D_+ \lesssim 2.6$ (upward arrows in Fig.
\ref{fig_v_z2468_c6_c8}a). A similar region is seen
in Fig. \ref{fig_v_z2468_c6_c8}b ($z_M>0$). The  downward arrows
in these figures indicate the location of such regions for HNC/MSA.
Therefore, the change in
sign of the derivative of the MEP observed in
Fig. \ref{zeta_all_c6_c8}b appears because the OHP is in a
LIDMEP region, where

\begin{equation}
\frac{\partial \psi(\sigma_0,r'=D_+)}{\partial \sigma_0} < 0.
\label{inlocpsi}
\end{equation}

Since ionic size and size asymmetry are evidently important
in  real highly coupled electrokinetic systems,
these last results suggest that the interpretation of the zeta potential
$\zeta$ in electrokinetic experiments must be done carefully
because the experimental MEP profiles could present LIDMEP regions,
according to our model calculations. Consequently,
the measured $\zeta$ might correspond to potentials
located in LIDMEP regions, giving rise to the possibility
of decreasing $\zeta$ for increasing $\sigma_0$ 
and vice versa.      
The occurrence of these non--monotonic
effects in the potential are clearly precluded in the PB-based
interpretation of the experimental measurements, as evidenced
in the present study.

\section{SUMMARY AND CONCLUSIONS}

In this paper the size-asymmetric electrical double layer (EDL) of
1:1 and 2:2 salts around a
slightly charged spherical macroparticle was studied
by assuming the primitive model of an electrolyte as a representation of
the ionic bath, and using
Monte Carlo simulations and the HNC/MSA integral equation in order
to calculate the corresponding properties of the model system.
Therefore, in the simulation and HNC/MSA integral equation the important ionic size
and size asymmetry effects have been fully incorporated through
the consistent consideration of the colloid-ions and ion-ion
interactions. Additionally, the possible consequences of a partial
treatment of such excluded volume and size asymmetry contributions
in the properties of the spherical EDL were assessed via the comparison with
the predictions of the classic URMGC theory for spherical geometry.
In the semi-punctual URMGC formalism the
finite nature of the ions is considered only through the use of two different
distances of closest approach between the colloid and counterions and colloid and coions.
Our Monte Carlo data evince that the finite size and size
asymmetry of the ionic species near
a barely charged colloid produce remarkable
phenomena, such as charge reversal, overcharging, the existence of potentials of zero charge,
and of a zero charge electrical double layer,
as well as the possibility of oscillations
in the ionic densities, and of sign alternacy in the mean electrostatic
potential. The simulations also show that the extent of all these
features is intimately linked to the ability of the smallest ions to
penetrate the Helmholtz zone.
Importantly, in this report we present the first simulational corroboration of overcharging
(i.e. of the increment of the native colloidal charge prompted by an anomalous
adsorption of coions), a peculiarity originally predicted
by Jim\'enez-\'Angeles et al. \cite{Ji01} in a more complex EDL system. Here it is
evidenced that, close to the point of zero charge, overcharging is a 
direct consequence of the size asymmetry of the ions.
On the whole, the HNC/MSA results coincide with the simulations, then providing an essentially correct
picture of the size-asymmetric SEDL near the point of zero charge. In contrast, URMGC disagrees quantitatively
and qualitatively with the Monte Carlo trends in most of the situations examined here.
Notably, and due to its inconsistent treatment of the
hard-core and electrostatic correlations, URMGC exhibits a spatial localization of all its ionic size and size asymmetry
effects within the zone delimited by the inner and outer Helmholtz planes.
One of the most characteristic sequels of the full incorporation of the ionic correlations in all the extension
of the EDL, and not only in the Helmholtz zone, is the occurrence of a non-monotonic behavior 
in the ionic concentration, charge density and mean electrostatic potential profiles
of Monte Carlo and HNC/MSA. In this respect, the existence of spatial regions
in which the integrated charge and mean electrostatic potential
present a non-uniform comportment with respect to the variation of the colloidal charge
proves its relevance for the understanding of the potential-charge relationship
at the Helmholtz planes. In particular, such ``non-uniform" regions serve to explain
the existence of a singular minimum in the $\psi_{OHP}-\sigma_0$ curve.

The plausible identification between the well-known zeta potential and the mean electrostatic potential around the Helmholtz
zone, which is a usual hypothesis in the
interpretation of electrophoretic mobility measurements, leads us to suggest several phenomena (e.g. the
presence of an anomalous sign in the zeta potential
and the occurrence of increased reversed mobilities)
which could be the objectives of future experimental protocols. As a first attempt in this direction, we
have pointed out the consistency between our results and some experimental data of
electrokinetic mobilities for $\alpha$-alumina particles. The eventual confirmation in the laboratory
of any of the theoretical predictions presented in this work (for example, charge reversal and overcharging) by means of
some electrokinetic or
static technique could indicate the pertinence of our simple primitive-model-based representation of
the size-asymmetric spherical EDL as a starting point to develop more faithful
descriptions of real colloidal systems, either in equilibrium or under external fields.
Finally, and in relation with this, the rich phenomenology discussed in the present communication, that arises from both the ionic size
and size asymmetry effects, puts a word of caution about the possible usage of the
Poisson-Boltzmann viewpoint and other contingent formalisms, such as the
so-called standard
electrokinetic model \cite{Bor01}, outside their range of applicability.

\begin{acknowledgments}
This work was supported by Consejo Nacional de Ciencia y Tecnolog\'{\i}a (CONACYT, M\'exico), 
through grants CB-2006-01/58470 and C01-47611, and PROMEP. E. G.-T. thanks CNS-IPICYT for 
computing facilities, and G. I. G.-G. acknowledges posdoctoral fellowships from CONACYT and 
Instituto Mexicano del Petr\'oleo. 
\end{acknowledgments}

 
\clearpage

\end{document}